\documentstyle[12pt]{article}        

\begin{document}               
\title{Birth and Early Development of Indian Astronomy}

\author{Subhash Kak\\
Louisiana State University\\
Baton Rouge, LA 70803-5901, USA}
\vspace{1in}
\date{In {\it Astronomy Across Cultures: The History of Non-Western Astronomy,}\\
Helaine Selin (ed), Kluwer, 2000, pp. 303-340}
\maketitle
\thispagestyle{empty}
\newpage

\section{Introduction}

In the last decade or so our understanding of the origin and
development of Indian astronomy and its relevance for
Indian religion and culture have undergone a major shift.
This shift has been caused by two factors: first, archaeological
discoveries that reveal to us that the the Sarasvati river,
the great river of the {\em \d{R}gvedic} times, dried up before
1900 BCE, suggesting that this ancient text must be at least as
old as that epoch; second, discovery of an astronomy in the
Vedic texts. The assignment of a date to the drying up of the
Sarasvati river has been a great aid to 
sorting the confusion regarding the chronology of
the Indian texts, but it could not have come before
an analysis of the
excavations of the Harappan towns and settlements of the 3rd millennium BCE.
On the other hand, the neglect of the astronomy of the Vedic texts
was caused by the inability of the philologists and Sanskritists
who studied these texts during the last 
two centuries to appreciate their scientific references.

Owing to the importance of the astronomy of the earliest
period for understanding the entire scientific tradition
in India, we will, in this essay, focus primarily on the
pre-{\em Siddh\={a}ntic} period before \={A}ryabha\d{t}a.
The subsequent history of Indian astronomy is 
well described by the {\em Siddh\={a}ntas} themselves
and by the many reviews that have appeared in the
published literature.

The fundamental idea pervading Indian thought from the
most ancient times is that of
equivalence or connection ({\em bandhu})
amongst the adhidaiva ({\em devas} or stars), adhibh\={u}ta
(beings), and adhy\={a}tma (spirit).
These connections, between the astronomical, the terrestrial, 
the physiological and the psychological, represent the constant
theme in the discourse of Indian texts.
These connections are usually stated in terms of vertical
relationships, representing a
recursive system; but they are also described horizontally across
hierarchies where they represent metaphoric or structural
parallels.
Most often, the relationship is defined in terms of numbers or other
characteristics. An example is the 360 bones of the infant---which later
fuse into the 206 bones of the adult---and the 360 days of the year.
Likewise, the tripartite division of the cosmos into earth,
space, and sky is reflected in the tripartite psychological
types.

Although the Vedic books speak often about astronomical phenomena,
it is only recently that the astronomical substratum of the Vedas
has been examined (Kak 1992-1999).
One can see a plausible basis behind many connections.
Research has shown that all life comes with its inner clocks.
Living organisms have rhythms that are matched to the periods of the
sun or the moon.
There are quite precise biological clocks of 24-hour (according to
the day), 24 hour 50 minutes (according to the lunar day since the moon
rises roughly 50 minutes later every day) or its half representing the
tides, 29.5 days (the period from one new moon to the next), and
the year.
Monthly rhythms, averaging 29.5 days, are reflected in the
reproductive cycles of many marine plants and those of animals.
The menstrual period is a synodic month and the average duration
of pregnancy is nine synodic months.
There are other biological periodicities of longer durations.
These connections need not be merely numerical.
In its most general form is the {\em Upani\d{s}adic} equation between
the self ({\em \={a}tman}) and the universe ({\em brahman}).

It is tempting to view
{\it jyoti\d{s}a}, the science of light and astronomy, as the fundamental paradigm
for the Vedic system of knowledge.
Jyoti\d{s}a is a term that connotes not only the light of the
outer world, but also the light of the inner landscape.
Astronomy is best described as nak\d{s}atra-vidy\={a}
of the {\em Ch\={a}ndogya Upani\d{s}ad}, but
because of its popularity we will also use jyoti\d{s}a in its narrow
meaning of astronomy.
As defining our place in the cosmos 
and as a means to understand the nature of time, astronomy is
obviously a most basic science.

That astronomy reveals that the periods of the heavenly bodies
are incommensurate might have led to the notion that true
knowledge lies beyond empirical apar\={a} knowledge.
On the other hand, it is equally likely that it was a
deep analysis of the nature of perception and the
paradox of relationship of the perceptor to the whole that was the 
basis of Vedic thought, and the incommensurability of the motions in
the sky was a confirmation of the insight that
knowledge is recursive.
This Vedic view of knowledge seems to have informed the earliest
hymns so it does not appear
to be feasible to answer the question of which
came first.
Neither can we now answer the question whether
jyoti\d{s}a as pure astronomy was a precursor to a jyoti\d{s}a that
included
astrology.

Analysis of texts reveals that much of Vedic mythology
is a symbolic telling of astronomical knowledge.
Astronomy was the royal science not only because it was the
basis for the order in nature, but also because the inner space of man,
viewed as a microcosm mirroring the universe, 
could be fathomed through its insights.

\subsection{Of Ceremonies, Festivals, Rites}
The importance of jyoti\d{s}a for agriculture and other
secular purposes are obvious and
so
we begin with a brief account of rites and festivals.
These 
ceremonies and rituals reveal
that there existed several traditions
of astronomical lore;
these variations are marked by
the different books of {\em \'{S}rautas\={u}tra}.
Such variation is perfectly in accord with an age when
astronomy was a living science with different scholars providing
different explanations.
Since our purpose is not to go into the details of the Vedic texts,
we will describe ceremonies and rites selectively. 

Different points in
the turning year were marked by
celebrations.
The year, beginning with the full moon in the month Ph\={a}lguna
(or Caitra), was divided into three four-monthly, {\it c\={a}turm\={a}sya},
sacrifices.
Another way of marking the year is by a year-long d\={\i}k\d{s}\={a}.
The year was closed with rites to
celebrate Indra \'{S}un\={a}\'{s}\={\i}ra
(Indra with the plough) to ``obtain the thirteenth month;''
this thirteenth month was interposed twice in
five years to bring the lunar year in harmony with the solar year.
This closing rite is to mark the first ploughing, in preparation for
the next year. Symbolically, this closing was taken to represent the
regeneration of the year.

Year-long ceremonies for the king's priest are described in the
{\em Atharvaveda} Pari\'{s}i\d{s}\d{t}a; these include those for the
health of horses, the safety of vehicles, and so on.
There existed other royal rites such as r\={a}jas\={u}ya,
v\={a}japeya and the a\'{s}vamedha, the so-called horse sacrifice,
which actually represented the transcendence by the king of time
in its metaphorical representation as horse.
The primary meaning of a\'{s}va as the sun is attested to in the
{\em \d{R}gveda}, {\em Nirukta}, and
{\em \'{S}atapatha Br\={a}hma\d{n}a}.

The {\em G\d{r}hyas\={u}tras} describe rites that mark the passage of the
day such as the daily agnihotra.
Three soma pressings, at sunrise, midday and sunset, were a part of
the daily ritual of agni\d{s}\d{t}oma.
Then there were the full and new moon ceremonies.
Longer soma rites were done as sattras, sessions of twelve days or
more.

\subsection{Altars}

Altar ritual was an important part of Vedic life and
we come across fire altars in the {\em \d{R}gvedic} hymns.
Study of Vedic ritual has shown that
the altar, adhiyaj\~{n}a, was used to show the
connections between the astronomical, the physiological
and the spiritual, symbolically.
That the altars represented
astronomical knowledge is what interests us in this article.
But the astronomy of the altars was not systematically
spelled out although there are pointed references in many texts including
the tenth chapter
of {\em \'{S}atapatha Br\={a}hma\d{n}a}, entitled ``Agnirahasya''.
{\em \d{R}gveda} itself is viewed as an 
altar of mantras in the {\em \'{S}ulbas\={u}tras}.

Altars were used in relation to
two basic types of Vedic ritual: \'{S}rauta and
G\d{r}hya.
This ritual marked specific points in the day or the year 
as in
the soma rituals of agni\d{s}\d{t}oma
and agnicayana.
The {\em \'{S}atapatha Br\={a}hma\d{n}a} 
describes
the twelve-day agnicayana rite that takes place in a large
trapezoidal area, called the mah\={a}vedi, and in a smaller
rectangular area to the west of it, which is called the
pr\={a}c\={\i}nava\d{m}\'{s}a
or pr\={a}gva\d{m}\'{s}a.
The text says clearly that agnicayana represents ritual as
well as knowledge.

The mah\={a}vedi trapezium measures 30 prakrama on the west,
24 prakrama on the east, and 36 prakrama lengthwise.
The choice of these numbers is related to the sum
of these three equaling one fourth the year or 90 days.

The nominal year of 360 days was used to reconcile the
discrepancies between the lunar and solar calendars, both of
which were used.
In the mah\={a}vedi a brick altar is built to represent time
in the form of a falcon about to take wing (Figure 1), and in the 
pr\={a}c\={\i}nava\d{m}\'{s}a
there are three fire altars in specified
positions, the g\={a}rhapatya, \={a}havan\={\i}ya, and
dak\d{s}i\d{n}\={a}gni.
The g\={a}rhapatya, which is round,
is the householder's fire received from the
father and transmitted to the descendents.
It is a perpetual fire from which the other fires are lighted.
The 
dak\d{s}i\d{n}\={a}gni is half-moon shaped; it
is also called the 
anv\={a}h\={a}ryapacana where cooking is done.
The \={a}havan\={\i}ya is square.
Between the \={a}havan\={\i}ya and the g\={a}rhapatya a space of
a rough hourglass is dug out and strewn with grass; this is called
the vedi and it is meant for the gods to sit on.

During the agnicayana ritual the old \={a}havan\={\i}ya serves
the function of the original g\={a}rhapatya.
This is the reason why their areas are to be identical, although
one of them is round and the other square.
In addition eight dhi\d{s}\d{n}ya hearths are built on an
expanded ritual ground.

Agnicayana altars are supposed to symbolize the universe.
G\={a}rhapa\-tya represents the earth, 
the dhi\d{s}\d{n}ya hearths represent space, and the
\={a}havan\={\i}ya altar represents sky.
This last altar is made in five layers.
The sky is taken to represent the universe therefore it includes space
and earth.
The first layer represents the earth, the third the space, and the fifth
the sky.
The second layer represents the joining of the earth and space,
whereas the fourth layer represents the joining of space and sky.

Time is represented by the metaphor of a bird.
The months of the year were ordinarily divided into six seasons
unless the metaphor of the bird for the year was used when
hemanta and \'{s}i\'{s}ira were lumped together.
The year as a bird had the head as vasanta, the body as
hemanta and \'{s}i\'{s}ira, the two wings as \'{s}arada
and gr\={\i}\d{s}ma, and the tail as var\d{s}\={a}.

The Vedic sacrifice is meant to capture the magic of change,
of time in motion.
Put differently, the altar ritual is meant to symbolize the paradoxes of
separation and unity, belonging and renunciation, and permanence
and death.
The yajam\={a}na, the patron at whose expense the ritual is
performed,
symbolically represents the universe.

The ritual culminates in his ritual rebirth, which signified the
regeneration of his universe.
In other words, the ritual is a play dealing with paradoxes of
life and death enacted for the yajam\={a}na's family and friends.
In this play symbolic deaths of animals and humans, including the
yajam\={a}na himself, may be enacted.

\subsection{Evolution of Vedic Thought}

How did the use of altars for a symbolic representation of knowledge
begin?
This development is described in the {\em Pur\={a}\d{n}as} where it is
claimed that the three altars were first devised by the king
Pur\={u}ravas.
The genealogical lists of the {\em Pur\={a}\d{n}as} and the epics provide
a framework in which the composition
of the different hymns can be seen.
The ideas can then be checked against social processes at work
as revealed by textual and archaeological data.

As we will see later in this article, there existed 
an astronomical basis to the organization of
the {\em \d{R}gveda} itself; this
helps us see Vedic ritual in a new light.
That astronomy could be used for fixing the chronology of certain
events in the Vedic books was
shown more than a hundred years ago by
Tilak and Jacobi.
This internal evidence compels the conclusion that the prehistory
of the Vedic people in India goes back to the fourth millennium
and earlier.
On the other hand, new archaeological
discoveries show a continuity in the Indian tradition going
as far back as 8000 BCE (Shaffer and Lichtenstein, 1995).
These are some of the elements in accord with the view that the
Vedic texts and the archaeological finds relate to the same reality.
One must also note that the rock art tradition in India has been
traced back to about 40000 BC (Wakankar, 1992). Whether this 
tradition gave birth to the Harappan tradition is not
clear at this time.

Recent archaeological discoveries establish that the Sarasvat\={\i} river
dried up around 1900 BCE which led to the collapse of the
Harappan civilization that was principally located in
the Sarasvat\={\i} region.
Francfort (1992) has even argued that the D\d{r}\d{s}advat\={\i} was already
dry before 2600 BCE.
The region of the Sarasvat\={\i} and the D\d{r}\d{s}advat\={\i}
rivers, called Brahm\={a}varta, was especially sanctified
and Sarasvat\={\i} was one of the mightiest
rivers of the {\em \d{R}gvedic} period.
On the other hand, Pa\~{n}cavi\d{m}\'{s}a Br\={a}hma\d{n}a describes
the disappearance of Sarasvat\={\i} in the sands at a distance of forty days on
horseback from its source.
With the understanding of the drying
up of Sarasvat\={\i} it follows that
the {\em \d{R}gvedic} hymns are generally
anterior to 1900 BCE but
if one accepts Francfort's interpretation of the data 
on the D\d{r}\d{s}advat\={\i} then the {\em \d{R}gvedic}
period includes the period before 2600 BCE.

It is most significant that the {\em Pur\={a}\d{n}ic}
king-lists speak of 1924 BCE as the epoch of the
{\em Mah\={a}bh\={a}rata} War, that marked the
end of the Vedic age. This figure of 1924 BCE emerges
from the count of 1500 years for the reigns prior to
the Nandas (424 BCE), quoted at several places
in the {\em Pur\={a}\d{n}as}.
Since this epoch is virtually
identical to the rough date of 1900 BCE for the
catastrophic drying up of the Sarasvati river, it
suggests that the two might have been linked if not
being the same, and it
increases our confidence in the use of the Indian texts
as sources of historical record.

\section{Nak\d{s}atras}

The {\em \d{R}gveda} describes the universe to be infinite.
Of the five planets it mentions B\d{r}haspati (Jupiter)
and Vena (Venus) by name
The moon's path was divided into 27 equal parts, although the moon takes
about 27 1/3 days to complete it.
Each of these parts was called a nak\d{s}atra.
A traditional iconic representation of the
nak\d{s}atras is shown in Figure 2.
Specific stars or asterisms were also termed
nak\d{s}atras, and they
are mentioned in the {\em \d{R}gveda} and {\em Taittir\={\i}ya
Sa\d{m}hit\={a}}, the latter specifically saying that they are
linked to the moon's path.
The {\em \d{R}gvedic} reference to 34 lights apparently means the sun, the moon,
the five planets, and the 27 nak\d{s}atras. 
In later literature the list of nak\d{s}atras was increased to 28.
Constellations other than the nak\d{s}atras were also known;
these include the \d{R}k\d{s}as (the Bears),
the two divine Dogs (Canis Major and Canis Minor), and the
Boat (Argo Navis).
{\em Aitareya Br\={a}hma\d{n}a} speaks of 
M\d{r}ga (Orion) and M\d{r}gavy\={a}dha (Sirius).
The moon is called s\={u}rya ra\'{s}mi, one that shines by
sunlight.

The constellations conjoined monthly with the circuit
of the sun were traditionally represented as in the
outer circle of Figure 3. The inner circle of this figure
shows the five planets, the sun, the moon and its
ascending and descending nodes.

The {\em \'{S}ata\-pa\-tha Br\={a}hma\d{n}a}
provides
an overview of the broad aspects of Vedic astronomy.
The sixth chapter of the book provides
significant clues.
Speaking of creation 
under the aegis of the Praj\={a}pati (reference either to a star
or to abstract time)
mention is made of the emergence of
A\'{s}va, R\={a}sabha, Aja and K\={u}rma before the emergence of the
earth.
It has been argued that these refer to stars or constellations.
Vi\'{s}van\={a}tha Vidy\={a}la\.{n}k\={a}ra (1985) suggests that these
should be identified as the sun (A\'{s}va),
Gemini (R\={a}sabha), Aja (Capricorn) and
K\={u}rma or 
K\={a}\'{s}yap\={\i}ya
(Cassiopeia).
This identification is supported by 
etymological considerations.
RV 1.164.2 and {\em Nirukta} 4.4.27 define A\'{s}va as the sun.
R\={a}sabha which literally means the twin asses 
are defined in {\em Nighan\d{t}u} 1.15 as A\'{s}vinau which later usage
suggests are 
Castor and Pollux in Gemini.
In Western astronomy the twin asses are to be found in the next
constellation of Cancer as Asellus Borealis and Asellus
Australis.
Aja (goat) is defined by Nighan\d{t}u 1.15 as a sun and owing to
the continuity that we see in the Vedic and later European names
for constellations (as in the case of the Great Bear) it is
reasonable to identify it as the constellation Capricorn
({\em caper} goat + {\em cornu} horn).

Vedic ritual was based on the times for the full and the new moons,
solstices and the equinoxes.
The year was known to be somewhat more than 365 days and a bit less
than 366 days.
The solar year was marked variously in the many different astronomical
traditions that marked the Vedic world.
In one tradition, an extra eleven days, marked by ek\={a}da\'{s}ar\={a}tra
or eleven-day sacrifice, were added to the lunar year of 354 days.
According to the {\em Taittir\={\i}ya Sa\d{m}hit\={a}} five more days are
required over the nominal year of 360 days to complete the seasons,
adding that four days are too short and six days are too long. 
In other traditions,
gav\={a}m ayana, `the walk of cows or intercalary periods,'
varied from 36 days of the lunar sidereal year of 12 months of 27 days,
to 9 days for the lunar sidereal year of 13 months of 27 days to
bring the year in line with the ideal year of 360 days; additional
days were required to be in accord with the solar year.

The year was divided into two halves: uttar\={a}yana, when the sun
travels north, and dak\d{s}i\d{n}\={a}yana, when the sun travels south.
According to the {\em Kau\d{s}\={\i}taki Br\={a}hma\d{n}a}, the year-long
sacrifices began with the winter solstice, noting the 
occurrence of the summer solstice,
vi\d{s}uvant, after six months.

The twelve tropical months, and the six seasons, are named in the
{\em Yajurveda}:
\begin{quote}
Madhu, M\={a}dhava in vasanta (spring);\\
\'{S}ukra, \'{S}uci in gr\={\i}\d{s}ma (summer);\\
Nabha, Nabhasya in var\d{s}\={a} (rains);\\
I\d{s}a, \={U}rja in \'{s}arada (autumn);\\
Saha, Sahasya in  hemanta (winter);\\
Tapa, Tapasya in \'{s}i\'{s}ira (freeze).\\
\end{quote}

The n\={a}k\d{s}atra names of the months began with Caitra in spring,
although some lists begin with Ph\={a}lguna. Since the months shift
with respect to the twelve nak\d{s}atra about 2,000 years per 
nak\d{s}atra, this change in the lists indicates a 
corresponding long period.
The lists that begin with Caitra mark the months thus:
\begin{quote}
Caitra, Vai\'{s}\={a}kha,\\
Jyai\d{s}\d{t}ha, \={A}\d{s}\={a}\d{d}ha,\\
\'{S}r\={a}va\d{n}a, Bh\={a}drapada,\\
\={A}\'{s}vina, K\={a}rttika,\\
M\={a}rga\'{s}ira, Pau\d{s}ya,\\
M\={a}gha, Ph\={a}lguna.\\
\end{quote}

The earliest lists of
nak\d{s}atras in the Vedic books begin with K\d{r}ttik\={a}s, the Pleiades;
much later lists dating from sixth century CE begin with A\'{s}vin\={\i}
when the vernal equinox occurred on the
border of Revat\={\i} and A\'{s}vin\={\i}.
Assuming that the beginning of the list marked the same astronomical
event, as is supported by other evidence, 
the earliest lists should belong to the third
millennium BCE.
The {\em Taittir\={\i}ya Sa\d{m}hit\={a}} 4.4.10 and
{\em \'{S}atapatha Br\={a}hma\d{n}a} 10.5.4.5 each mention 27 nak\d{s}atras.
But there was also a tradition of the use of 28 nak\d{s}atras.
The {\em Atharvaveda} 19.7
lists these 28 together with their presiding deities; the
additional nak\d{s}atra is Abhijit.
The lists begins with
K\d{r}ttik\={a} (Pleiades) where the spring
equinox was situated at that time.

\subsection{Nak\d{s}atras and chronology}

Motivated by the then-current models of the movements of pre-historic
peoples, it became, by the end of the nineteenth century,
fashionable in Indological
circles to dismiss any early astronomical references
in the Vedic literature.
But since the publication of {\it Hamlet's Mill: An essay on myth and
the frame of time} by Georgio de Santillana and Hertha von Dechend
in 1969 it has come to be generally recognized that ancient myths
encode a vast and complex body of astronomical knowledge.
The cross-checks provided by the dating of some of the Indian myths
provide confirmation to the explicit astronomical evidence related to
the nak\d{s}atras that is spelled out below.
Other confirmation comes from the archaeological evidence summarized
in this article.

Due to the precession of the earth's polar axis the direction of
the north pole with respect to the fixed background stars keeps on
changing.  The period of this precession is roughly 26,000.  Polaris
($\alpha$ Ursae Minoris) is the Pole star now but around 3000 BCE it was
$\alpha$ Draconis which was followed later by $\beta$ Ursae Minoris; in CE
14000 it will be Vega.
The equinoxes and the solstices also shift with
respect to the background stars.
The equinoxes move along the ecliptic in a
direction opposite to the yearly course of the sun (Taurus to Aries to Pisces
rather than Pisces to Aries to Taurus and so on).

The vernal equinox marked an important day in the year.
The sun's position among the constellations at the vernal equinox
was an indication of the state of the precessional cycle.  This
constellation was noted by its heliacal rising.  The equinoctial sun
occupies each zodiacal constellation for about 2200 years.  Around 5000
BCE it was in Gemini; it has moved since into Taurus, Aries, and is now
in Pisces.
The sun spends about 13 1/3 days in each nak\d{s}atra, and the
precession of the equinoxes takes them across each nak\d{s}atra in about a
1000 years.

Thirteen and a half nak\d{s}atras ending with Vi\'{s}\={a}kh\={a}
were situated in the
northern hemispheres; these were called devanak\d{s}atras.
The remaining
nak\d{s}atras
ending with Bhara\d{n}\={\i} that were in the southern hemisphere were
called yamanak\d{s}atras (yama: twin, dual).
This classification in
the {\em Taittir\={\i}ya Br\={a}hma\d{n}a} (1.5.2.7) corresponds
to 2300 BCE.

As mentioned above, the list beginning with K\d{r}ttik\={a} indicates
that it was drawn up in the third millennium BCE.
The legend of the cutting off of Praj\={a}pati's head suggests
a time when the year began with M\d{r}ga\'{s}\={\i}r\d{s}a
in the fifth millennium BCE.
Scholars have also argued that a subsequent list began with
Rohi\d{n}\={\i}. This view is strengthened by the fact that there are
two Rohi\d{n}\={\i}s, separated by fourteen nak\d{s}atras,
indicating that the two 
marked the beginning of the two half-years.

The {\em \'{S}atapatha Br\={a}hma\d{n}a} speaks of a marriage between the 
Seven Sages, the stars of the Ursa Major, and the K\d{r}ttik\={a}s;
this is elaborated in the {\em Pur\={a}\d{n}as} where it is stated that the
\d{r}\d{s}is remain for a hundred years in each nak\d{s}atra.
In other words, during the earliest times in India there existed a
centennial calendar with a cycle of 2,700 years.
Called the Saptar\d{s}i calendar, 
it is still in use in several parts of India. Its current beginning is taken to
be 3076 BCE. 
On the other hand, notices by the Greek historians Pliny and
Arrian suggest that, during the 
Mauryan times, the calendar used in India began in 6676 BCE.
It is very likely that this calendar was the Saptar\d{s}i calendar with
a beginning at 6676 BCE.

Around 500 CE, a major review of the Indian calendar was attempted
by astronomers. \={A}ryabha\d{t}a, Var\={a}hamihira and others used
the nak\d{s}atra references that the
Saptar\d{s}i were in Magh\={a} at the time of the
{\em Mah\={a}bh\={a}rata} war to determine its epoch.
\={A}ryabha\d{t}a declared the war to have occurred in 3137 BCE
(the Kaliyuga era begins 35 years after the war), and 
Var\={a}hamihira assigned it 2449 BCE.
It has been suggested that this discrepancy arose because 
the change in the number of nak\d{s}atras from the earlier counts of
27 to the later 28 was differently computed by the two astronomers.
It is quite likely that the fame of the Kaliyuga era with its
beginning assigned to 3102 BCE prompted a change in the beginning
of the Saptar\d{s}i era to about the same time, viz. to 3076
BCE.

The shifting of seasons through the year and the shifting of the northern
axis allow us to date several other statements in the books.
Thus
the {\em \'{S}atapatha Br\={a}hma\d{n}a} (2.1.2.3) has a statement that points to
an earlier epoch where it is stated that K\d{r}ttik\={a} never swerve from the
east.
This correspond to 2950 BCE.

The {\em Maitray\={a}n\={\i}ya Br\={a}hma\d{n}a Upani\d{s}ad}
(6.14) refers to the winter
solstice being at the mid-point of the 
\'{S}ravi\d{s}\d{t}h\={a} segment and the
summer solstice at the beginning of Magh\={a}.
This indicates 1660 BCE.

The {\em Ved\={a}\.{n}ga Jyoti\d{s}a} (Yajur 6-8) mentions that winter solstice was at
the beginning of 
\'{S}ravi\d{s}\d{t}h\={a} 
and the summer solstice at the mid-point of
A\'{s}le\d{s}\={a}.  This corresponds to about 1370 BCE (Sastry, 1985).

It should be noted that these dates can only be considered to be
very approximate. Furthermore, these dates do not imply that the
texts come from the corresponding period; the text may recall
an old tradition.
A chronology of the Vedic period by
means of astronomical references was attempted by the
historian of science P.C. Sengupta.
Amongst other evidence, Sengupta uses the description of the
solar eclipse in RV 5.40.5-9 to fix a date for it.
Unfortunately, this work has not received the attention
it deserves.

The changes
in the beginning of the Nak\d{s}atra lists
bring us down to the
Common Era;
at the time of Var\={a}hamihira (550 CE) the vernal equinox was
in A\'{s}vin\'{\i}.

\section{Ritual, geometry and astronomy}

We have mentioned that
the altars used in the ritual were based on astronomical
numbers related to the reconciliation of the lunar and solar
years.
The fire altars symbolized the universe
and there were three types of altars representing the 
earth, the space and the sky.
The altar for the earth was drawn as circular whereas the sky (or heaven)
altar was drawn as square.
The geometric problems of circulature of a square and that of
squaring a circle are a result of equating the earth and the sky
altars.
As we know these problems are among the earliest considered in ancient
geometry.

The fire altars were surrounded by 360 enclosing stones,
of these 21 were
around the earth altar, 78 around the space altar and 261
around the sky altar.
In other words, the earth, the space, and the sky are symbolically
assigned the numbers 21, 78, and 261.
Considering the earth/cosmos dichotomy, the two numbers are
21 and 339 since cosmos includes the space and the sky.

The main altar was built in five layers. The basic square shape
was modified to several forms, such as falcon and turtle.
These altars were built in five layers, of a thousand bricks of
specified shapes.
The construction of these altars required the solution to several
geometric and algebraic problems.

Two different kinds of bricks were used: the special and the
ordinary.
The total number of the special bricks used was 396,
explained as 360 days of the year and the additional
36 days of the intercalary month.
By layers, the first has 98, the second has 41, the third has 71,
the fourth has 47 and the fifth has 138.
The sum of the bricks in the fourth and the fifth layers equals
186 tithis of the half-year.
The number of bricks in the third and the fourth layers equals the
integer nearest to one third the number of days in the lunar year,
and the number of bricks in the third layer equals the 
integer nearest to one fifth of the number of days in the lunar
year, and so on.

The number of ordinary bricks equals 10,800 which equals the
number of muh\={u}rtas
in a year (1 day = 30 muh\={u}rtas), or equivalently the number of
days in 30 years. Of these 21 go into the g\={a}rhapatya,
78 into the eight dhi\d{s}\d{n}ya hearths, and the rest go
into the \={a}havan\={\i}ya altar.

\subsection{Equivalence by area}
The main altar was an area of 7$\frac{1}{2}$ units. This area was taken to
be equivalent to the nominal year of 360 days.
Now, each subsequent year, the shape was to be reproduced with
the area increased by one unit.

The ancient Indians spoke of two kinds of
day counts: the solar day, and tithi, 
whose mean value is the lunar year divided into
360 parts. They also considered three different years: (1) nak\d{s}atra,
or a year of 324 days (sometimes 324 tithis)
obtained by considering 12 months of 27 days each, where
this 27 is the ideal number of days in a lunar month;
(2) lunar, which is a fraction more than 354 days (360 tithis); and (3) solar,
which is in excess of 365 days (between 371 and 372 tithis).
A well-known altar ritual says that altars should be constructed
in a sequence of 95, with progressively increasing areas.
The increase in the area, by one unit yearly, in building progressively
larger fire altars is 48 tithis which is about equal to
the intercalation required to make the nak\d{s}atra year in tithis equal to the
solar year in tithis.
But there is a residual excess which in
95
years adds up to 89 tithis; it appears that after this
period such a correction was made.
The 95 year cycle corresponds to the
tropical year being equal to 365.24675 days.
The cycles needed to harmonize various motions led to the concept
of increasing periods and world ages.

\subsection{The \d{R}gvedic altar}
The number of syllables in the
{\it \d{R}gveda} confirms the textual references that the book was
to represent a symbolic
altar.
According to various early texts, the number of syllables in the
{\em \d{R}gveda} is 432,000, which is the number of
muh\={u}rtas
in forty years.
In reality the syllable count is somewhat less because certain
syllables are supposed to be left unspoken.

The verse count of the {\em \d{R}gveda} can be
viewed as the number of sky days in forty years or
$261 \times 40 = 10,440$, and the
verse count of all the Vedas is $261 \times 78 = 20,358 $.

The {\em \d{R}gveda} is divided into ten books with a total of 1,017 hymns which
are placed into 216 groups. 
Are these numbers accidental or is there a deliberate plan behind the
choice?
One would expect that if the {\em \d{R}gveda} is considered akin to the 
five-layered altar described in the Br\={a}hma\d{n}as
then the first two books should correspond to the space
intermediate to the earth and the sky.
Now the number that represents space is 78.
When used with the multiplier of 3 for the three worlds, this
yields a total of 234 hymns
which is indeed the number of hymns in these two books.
One may represent the {\em \d{R}gvedic} books as a five-layered altar of
books as shown in Table 1.

Table 1: The altar of books

\begin{tabular}{||c|c||}  \hline
Book 10 &	Book 9 \\
Book 7 &	Book 8 \\
Book 5 &	Book 6 \\
Book 3 &	Book 4 \\
Book 2 &	Book 1 \\ \hline
\end{tabular}

When the hymn numbers are used in this altar of books we obtain Table 2.

Table 2: Hymns in the altar of books 

\begin{tabular}{||r|r||} \hline
191 &	114 \\
104 &	92 \\
87 &	75 \\
62 &	58 \\
43 &	191 \\ \hline
\end{tabular}

\par
 
The choice of this arrangement is prompted by the considerable
regularity in the hymn counts.
Thus the hymn count separations diagonally across the two columns
are 29 each for Book 4 to Book 5 and Book 6 to Book 7 and they are
17 each for the second column for Book 4 to Book 6 and Book 6 to Book 8.
Books 5 and 7 in the first column are also separated by 17;
Books 5 and 7 also add up to the total for either Book 1
or Book 10.
Another regularity is that the middle three layers are indexed by
order from left to right whereas the bottom and the top layers are
in the opposite sequence.
 
Furthermore, Books [4+6+8+9] = 339, and these books may be taken to
represent the spine of the altar.
The underside of the altar now consists of the
Books [2+3+5+7] = 296, and the feet and the head Books [1+10]
= 382.
The numbers 296 and 382 are each 43 removed from the fundamental
{\em \d{R}gvedic} number of 339.

The {\em Br\={a}hma\d{n}as} and the {\em \'{S}ulbas\={u}tra} tell us about
the altar of chandas and meters, so we would expect that
the total hymn count of 1017 and
the group count of 216 have particular significance.
Owing to the pervasive tripartite ideology of the
Vedic books we choose to view
the hymn number as $339 \times 3$.
The tripartite ideology refers to the consideration of time in
three divisions of past, present, and future and the consideration of
space in the three divisions of the northern celestial hemisphere,
the plane that is at right angle to the earth's axis, and the
southern celestial hemisphere.
 
Consider the two numbers
1017 and 216.
One can argue that another parallel with the representation of the
layered altar was at work in the group total of 216.
Since the {\em \d{R}gvedic} altar of hymns was meant to symbolically
take one to the sky, the abode of gods, it appears that the
number 216 represents twice the basic distance of 108 taken
to separate the earth from the sky.
The {\em \d{R}gvedic} code then expresses a fundamental connection between
the numbers 339 and 108.
 
Consider now the cosmic model used by the ancients.
The earth is at the center, and the sun and the moon orbit the
earth at different distances.
If the number 108 was taken to represent symbolically the
distance between the earth and the sky, the question arises
as to why it was done.
The answer is apparent if one considers the actual distances of the
sun and the moon.
The number 108
is roughly the average distance that the sun is in terms of
its own diameter from the earth; likewise, it is also the average
distance that the moon is in terms of its own diameter from the 
earth. 
It is owing to this marvellous coincidence that the angular size of
the sun and the moon, viewed from the earth, is about identical.
 
It is easy to compute this number.
The angular measurement of the sun can be obtained quite easily during
an eclipse.
The angular measurement of the moon can be made on any clear full moon
night.
A easy check on this measurement would be to make a person hold a pole
at a distance that is exactly 108 times its length and confirm that
the angular measurement is the same.
Nevertheless, the computation of this number would require careful
observations.
Note that 108 is an average and due to the ellipticity of the
orbits of the earth and the moon the distances vary with the seasons.
It is likely, therefore, that observations did not lead to the precise number
108, but it was chosen as the true value of the distance since it is
equal to
$ 27 \times 4 $, because of the mapping of the sky into
27 nak\d{s}atras.
 
The second number 339 is simply the number of disks of the sun or the 
moon to measure the path across the sky:
$\pi \times 108 \approx 339. $

We return to a further examination of the numbers
296, 339, and 382 in the design of the {\em \d{R}gvedic} altar.
It has been suggested that 339 has an obvious significance as the
number of sun-steps during the average day or the equinox, and
the other numbers are likely to have a similar significance.
In other words, 296 is the number of sun-steps during the
winter solstice and 382 is the number of sun-steps during the
summer solstice (Kak, 1994).

There also exists compelling evidence, of a probabilistic sense, that the
periods of the 
planets had been obtained and used in the setting up of the
{\em \d{R}gvedic} astronomical code.

\section{The motions of the sun and the moon}

The {\em Ved\={a}\.{n}ga Jyoti\d{s}a} (VJ), due to Lagadha, is a 
text that describes some of the astronomical knowledge of the times of 
altar ritual. It
has an internal date of c. 1350 BCE
obtained
from its assertion that the winter solstice was at the 
asterism
\'{S}ravi\d{s}\d{t}h\={a}
(Delphini).
Recent archaeological discoveries 
support 
such an early date, and so this
book assumes great importance in the understanding of the
earliest astronomy.

VJ
describes the
mean motions of the sun and the moon.
This manual is available in two recensions: the earlier {\em \d{R}gvedic} VJ (RVJ)
and the later {\em Yajurvedic} VJ (YVJ).
RVJ has 36 verses and YVJ has 43 verses.
As the only extant astronomical text from the Vedic period, we describe
its contents in some detail.

The measures of time used in VJ are as follows:
\begin{quote}
1 lunar year = 360 tithis\\
1 solar year = 366 solar days\\
1 day = 30 muh\={u}rtas\\
1 muh\={u}rta = 2 n\={a}\d{d}ik\={a}s\\
1 n\={a}\d{d}ik\={a} = 10$\frac{1}{20}$  kal\={a}s\\
1 day = 124 a\.{m}\'{s}as (parts) \\
1 day = 603 kal\={a}s\\
\end{quote}

Furthermore, five years were taken to equal a yuga.
A ordinary yuga consisted of 1,830 days.
An intercalary month was added at half the yuga and another
at the end of the yuga.

What are the reasons for the use of a time division of the day into
603 kal\={a}s?
This is explained by the assertion that the moon travels through
1,809 nak\d{s}atras in a yuga.
Thus the moon travels through one nak\d{s}atra in 1$\frac{7}{603}$
sidereal days because 
\[1,809 \times 1 \frac{7}{603} = 1,830.\]

Or the moon travels through one nak\d{s}atra in 610 kal\={a}s.
Also note that 603 has 67, the number of sidereal months in a yuga,
as a factor
The further division of a kal\={a} into 124 k\={a}\d{s}\d{t}h\={a}s
was in symmetry with the division of a yuga into 62 synodic months
or 124 fortnights (of 15 tithis),
or parvans. A parvan is the angular distance travelled by
the sun from a full moon to a new moon or vice versa.

The VJ system is a coordinate system for the sun and the moon in
terms of the 27 nak\d{s}atras.
Several rules are given so that a specific tithi and nak\d{s}atra
can be readily computed.

\begin{quote}
The number of risings of the asterism \'{S}ravi\d{s}\d{t}h\={a}
in the yuga is the number of days plus five (1830+5 = 1835). The 
number of risings of the moon is the days minus 62
(1830-62 = 1768). The total of each of the moon's 27 asterisms
coming around 67 times in the yuga equals the number of days
minus 21 (1830-21 = 1809). \\

The moon is conjoined with each asterism 67 times during a yuga.
The sun stays in each asterism 13$\frac{5}{9}$ days.
\end{quote}

The explanations are straightforward. The sidereal risings equals the 
1,830 days together with the five solar cycles.
The lunar cycles equal the 62 synodic months plus the five solar cycles.
The moon's risings equal the risings of 
\'{S}ravi\d{s}\d{t}h\={a}
minus the moon's cycles.

This indicates that the moon was taken to rise at a mean rate of

$\frac{1,830}{1,768}$ = 24~hours~and~50.4864~minutes.

\subsection{Computation of tithis, nak\d{s}atras, kal\={a}s}
Although a mean tithi is obtained by considering the lunar
year to equal 360 tithis, the determination of a tithi each day
is by a calculation of a shift of the moon by 12$^\circ$ with respect to the
sun. In other words, in 30 tithis it will cover the full circle of
360$^\circ$. 
But the shift of 12$^\circ$ is in an irregular manner and
the duration of the tithi can vary from day to day. As a practical
method a mean tithi is defined by a formula.
VJ takes it to be 122 parts of the day
divided into 124 parts.

Each yuga was taken to begin with the asterism
\'{S}ravi\d{s}\d{t}h\={a}
and the synodic month of M\={a}gha, the solar month Tapas and the
bright fortnight (parvan), and the northward course of the sun and
the moon.
The intercalary months were used in a yuga. But since the civil year 
was 366 days, or 372 tithis, it was necessary to do further corrections.
As shown in the earlier section, a further correction was performed
at 95 year, perhaps at multiples of 19 years.

The day of the lunar month corresponds to the tithi at sunrise.
A tithi can be lost whenever it begins and ends between one sunrise
and the next.
Thus using such a mean system, the days of the month can vary in length.

\subsection{Accuracy}

There are other rules of a similar nature which are based on the
use of congruences.
These include rules on hour angle of nak\d{s}atras, time of the day at
the end of a tithi, time at the beginning of a nak\d{s}atra,
correction for the sidereal day, and so on.
But it is clear that the use of mean motions can lead to 
discrepancies that need to be corrected at the end of the yuga.

The framework of VJ has approximations built into it
such as consideration of the civil year to be 366 days and
the consideration of a tithi as being equal to $\frac{122}{124}$ of a day.
The error between the modern value of tithi and its VJ value is:
\[\frac{354.367}{360} - \frac{122}{124}\]
which is as small as $5 \times 10^{-4}$. This leads to an
error of less than a day in a yuga of five years.

The constructions of the geometric altars as well as the Vedic books
that come centuries before VJ 
confirm that the Vedic Indians knew
that the year was more than 365 days and less than 366 days.
The five year period of 1,830 days, rather than the more
accurate 1,826 days, was chosen because it is divisible by
61. This choice defines a symmetry with the definition of the
tithi as $\frac{61}{62}$ of the day.
The VJ system was thus very accurate for the
motions of the moon but it could have only served as a framework
for the motions of the sun.
It appears that
there were other rules of missing days that 
brought the calendar into consonance with the reality of the
nak\d{s}atras at the end of the five year yuga and at the end
of the 95 year cycle of altar construction.

Mean motion astronomy can lead to significant discrepancy between
true and computed values.
The system of intercalary months introduced further
irregularity into the system.
This means that the conjunction between the sun and the moon that
was assumed at the beginning of each yuga became more and more
out of joint until such time that the major extra-yuga corrections
were made.

Since the Vedic astronomers were evidently aware of the many corrections
that is required in the calendric system of the VJ, one might wonder about
the choice of its constants.
It appears that the yuga of 1,830 days, rather than the more accurate
1,826 days, was chosen because it is divisible by 61; this choice
simplifies computations for a tithi defined as $\frac{61}{62}$ of the
day.

\section{The planets}
Although it is certain that the planets had been studied by  the 
{\em \d{R}gvedic} people, we do not find a single place in the texts where the
names are listed together. 
The list below brings together some of the names, together with
the ascribed colours, used in a variety of places
including the later {\em Pur\={a}\d{n}ic} literature.

\begin{description}
\item MERCURY. Budha, Saumya, Rauhi\d{n}eya, Tu\.{n}ga ({\it yellow})
\item VENUS. Vena, U\'{s}anas, \'{S}ukra, Kavi, Bh\d{r}gu ({\it white })
\item MARS. A\.{n}g\={a}raka, Bh\={u}mija, Lohit\={a}\.{n}ga, Bhauma, Ma\.{n}gala, Kum\={a}ra, Skanda ({\it red })
\item JUPITER. B\d{r}haspati, Guru, \={A}\.{n}giras ({\it yellow})
\item SATURN. \'{S}anai\'{s}cara, Sauri, Manda, Pa\.{n}gu, P\={a}ta\.{n}gi
({\it black })
\end{description}

Mercury is viewed as the son of the moon by T\={a}r\={a}, the wife
of Jupiter, or the
nak\d{s}atra Rohi\d{n}i (Aldebaran), Venus as the son of Bh\d{r}gu and the priest of the
demons, Mars as the son of the earth or \'{S}iva, 
Jupiter as the son of A\.{n}giras and the priest of the gods,
and Saturn is seen as being born to Revat\={i} and Balar\={a}ma or
to Ch\={a}y\={a} and the sun.
Saturn is
described as
the lord of the planets, lord of seven lights or satellites, 
and the slow-goer.
Since the Indian calendar was reckoned according to the constellation at
the vernal equinox,
one may assume the name son of Aldebaran implies that Mercury was first
noted during the era of 3400-2210 BCE when the vernal equinox was in
the Pleiades.

The {\em Jaiminig\d{r}hyas\={u}tra} gives the following equation between
the planets and the Vedic gods: the sun is \'{S}iva; the moon is Um\={a} (\'{S}iva's
wife); Mars is Skanda, the son of \'{S}iva; Mercury is Vi\d{s}\d{n}u; Jupiter is
Brahman (symbolizing the entire universe); Venus is Indra; and Saturn
is Yama, the ``dual'' god (death). The colors assigned to the planets are
from the same source.

One may speculate that the equation of Saturn and Yama arises out of the
fact that the synodic period of Saturn is the ``dual'' to the lunar year;
378 days of Saturn and 354 days of the lunar year with the centre
at the 366-day solar
year.

\subsection{On the identity of Mercury and Vi\d{s}\d{n}u}

Mercury's identification with the god Vi\d{s}\d{n}u, an
important figure in the {\em \d{R}gveda}, is of particular significance.
Vi\d{s}\d{n}u is the youger brother of Indra in the {\em \d{R}gvedic} era;
and Indra is sometimes identified with the sun.
The most essential feature of Vi\d{s}\d{n}u are his three steps by which he
measures out the universe (e.g. RV 1.154).
Two of these steps are visible to men, but the third or highest step
is beyond the flight of birds or mortals (RV 1.155, 7.99).
In later mythology it is explained that Vi\d{s}\d{n}u did this remarkable thing
in the incarnation as V\={a}mana, the pygmy.
This agrees with the identification as the small Mercury.

Now what do these steps mean? According to late
tradition, Vi\d{s}\d{n}u is a solar
deity and so these three steps represent
the sunrise, the highest ascent, and the sunset.
Another equally old interpretation is that the three steps represent
the course of the sun through the three divisions of the universe:
heavens, earth, and the netherworld.

But both of these interpretations appear unsatisfactory.
Neither of these interpretations squares with the special significance
attached to the third step.
Nor does not explain the putative identity of Mercury and Vi\d{s}\d{n}u. 

An explanation becomes obvious when we consider the
Vedic altar ritual.
It appears likely that the three steps of
Vi\d{s}\d{n}u are nothing but the three revolutions of Mercury in
a cycle of 261 sky days.
With this supposition the period of Mercury will be 87 days.
Furthermore, three synodic periods of Mercury, at 
118 days a period,
equal the 354 lunar days or 360 tithis.
It appears that this dual relationship led to the great importance being
given to the myth of the three steps of Vi\d{s}\d{n}u.
Of course, the figures for the periods are only approximate but as expected
at the first determination of these numbers an attempt was made to
connect them to the basic numbers of 261 and 354.

The explicit name of Budha for Mercury appears in 
the {\em Pa\~{n}cavi\d{m}\'{s}a Br\={a}hma\d{n}a} (PB)
which is dated definitely after 1900 BCE since it has an account of
a journey to the source of Sarasvat\={i} from the place where it is
lost in the desert (PB 25.10).
PB 24.18 speaks of Budha in connection with a 61 day rite.
Three such rites imply a total of 183 days which equals the days exclusively
devoted to the heavens.
This appears to be the analog, in the field of ritual, of the three steps  of
Vi\d{s}\d{n}u covering the heavens.

We note that the understanding of the
motions of the planets arose at some time during the unfolding of the
{\em \d{R}gvedic} period.
For example, Venus is described in early Vedic mythology in
terms of the twin
A\'{s}vins, the morning and evening stars just
as Homer later describes it as the pair Hesperus and Phosphorus.
This commonality indicates early Indo-European basis to this myth.

The main characters in the planetary myths are Jupiter and Venus as is to
be expected for the two brightest planets.
Venus, in its earlier incarnation as the A\'{s}vin twins, was seen as born to
the sun.
Mercury as Vi\d{s}\d{n}u is Upendra, the younger brother of the Indra, here a
personification of the sun. But once Mercury fitted into the planetary
scheme, its association with Vi\d{s}\d{n}u was forgotten.
Later acccounts describe the planets in relation to each other.
Our arguments showing that the period of Mercury was
obtained in the third millennium BCE imply that
as the determination of the period of Mercury is the hardest amongst
the classical planets, 
the periods of the other planets had been
obtained.

The literature that followed the {\em \d{R}gvedic} age was at first concerned more with
the ritual related to the earlier astronomy of the Vedic age.
Once the planetary system fell into place, the gods became supernumeraries.
Now the focus shifted to their duals that inhabit the inner universe.
Thus by the time of the {\em \'{S}atapatha Br\={a}hma\d{n}a} (second millennium BCE),
the original stars of the Ursa Major were identified with the cognitive
centres in the brain as in \'{S}B 8.1 or in more detail in
BU 2.2.4.

The {\em \d{R}gveda} and the {\em \'{S}atapatha Br\={a}hma\d{n}a} speak of
the five planets as gods. There is also a mention of the
thirty-four lights, which appear to be the twenty seven
nak\d{s}atras, the five planets, the sun and the moon.
The moon is the fastest moving of the heavenly bodies, and so
it is compared to the male who activates or fertilizes the
other heavenly bodies with which it comes in contact.
The {\em \d{R}gveda} speaks of the five bulls of heaven, which appear
to be the five planets.
Being faster than the fixed stars, the planets can, in turn,
be compared to bulls.

The {\em Taittir\={\i}ya Sa\d{m}hit\={a}} speaks of the 33 daughters
of Praj\={a}pati, personification of time here, that are given
in marriage to Soma, the moon, viewed as king.
These are the 27 nak\d{s}atras, the five planets, and the sun.
The sun as the bride, S\={u}ry\={a}, is described in the {\em \d{R}gveda} and
the {\em Atharvaveda}.

Since the planets move through the nak\d{s}atras
and Venus and Jupiter are brighter than any of the stars, 
observation of the nak\d{s}atras presupposes a notice of the 
planets. The {\em Ved\={a}\.{n}ga Jyoti\d{s}a} does not mention the
planets, but that is so because its concern is only the motions of
the sun and the moon related to fixing the calendar. 

The rivalry 
between the families of A\.{n}girases
and the Bh\d{r}gus, mythical figures in the \d{R}gveda, represents
the motions of Jupiter and Venus. This is clear in later accounts
where B\d{r}haspati (Jupiter), the priest of the gods
because its motion is closest to the ecliptic, is an
\={A}\.{n}giras and Kavi U\'{s}anas or \'{S}ukra (Venus), a
Bh\={a}rgava, is the priest of the Asuras.

The idea of eclipse was expressed by the notion of R\={a}hu
seizing the heavenly body.
The fact that graha, `seize,' is the name used for planets right
from the time of {\em Atharvaveda} suggests that the 
waxing and waning of the two inferior planets, Mercury and Venus, as
well as the change in the intensity of the others was known.

Although there is mention of a week of six days, called a
\d{s}a\d{d}aha, in the early books, it does not follow that
the tradition of a week of seven days is a later one.
The seven day week was in use during the time of {\em Atharva Jyoti\d{s}a.}

The sidereal periods suggested by the astronomical code in the
organization of the {\em \d{R}gveda} are (Kak, 1994):
\begin{quote}
Mercury:~~87 days\\
Venus:~~225 days\\
Mars:~~687 days\\
Jupiter:~~4,340 or 4,350 days\\
Saturn:~~10,816 days.\\
\end{quote}

\subsection{Soma}
Soma, or the moon, is one of the most important deities of the {\em \d{R}gveda}.
It is related to S\={u}rya the way puru\d{s}a is related to
prak\d{r}ti.
Soma is almost always the moon in the ninth book of the
{\em \d{R}gveda}.
That very few Western scholars of the nineteenth century recognized
this fact can only be explained by recalling the incorrect assumptions
they labored under.
Soma, as a drink, was meant to celebrate the creative function of
the moon as reflected in the tides, the menstrual cycle and the
growth of plants.

\section{The Yuga concept}

There are allusions to yugas, meant as an age, in the Vedas.
In the {\em Aitareya Br\={a}hma\d{n}a}
Kali, Dv\={a}para, Tret\={a}, and K\d{r}ta are compared to a man
lying down, moving, rising, and walking. The {\em \d{S}a\d{d}vi\d{m}\'{s}a
Br\={a}hma\d{n}a} mentions the four ages Pu\d{s}ya, Dv\={a}para,
Kh\={a}rv\={a}, and K\d{r}ta.
In order from K\d{r}ta to Kali, each yuga represents a decline in
morality, piety, strength, knowledge, truthfulness, and happiness.
The notion of a yuga appears to have a historical basis.
If we accept that
a catastrophic tectonic event took place
around 1900 BCE, leading eventually
to a great shift in the population away from the
Sarasvat\={\i} valleys, then
Kaliyuga could be a memory of the beginning of that dark age.
Support for this view comes from the {\em Mah\={a}bharata}, according
to which all places were sacred in the K\d{r}tayuga;
Pu\d{s}kara in the Sarasvat\={\i} region was the
most sacred in Tret\={a}yuga;
Kuruk\d{s}etra in Dv\={a}para; and Pray\={a}ga at the junction of
Ga\.{n}g\={a} and Yamuna in the Kaliyuga. This clearly marks the
shift in focus of the Vedic people.

The five years of the yuga of the
{\em Ved\={a}\.{n}ga Jyoti\d{s}a }
are named variously; one text calling them
sa\d{m}vatsara, parivatsara, id\={a}vatsara, iduvatsara, and
vatsara.
It has been suggested that the 33 gods mentioned at many places
refer to a cycle of 33 years but this cannot be accepted until
corroborative evidence is found.
As mentioned
before, a cycle of 95 years is described in 
the {\em \'{S}atapatha Br\={a}hma\d{n}a.}
The yuga of 60 years appears to  have emerged out of an attempt to
harmonize the approximate sidereal periods of 12 and 30 years for
Jupiter and Saturn, respectively.
Consideration of more accurate sidereal values requires much larger
periods that are seen in the later {\em Siddh\={a}ntic} astronomy of the 
Classical period.

The {\em Pur\={a}\d{n}as} talk of a kalpa, a day of Brahm\={a} which is taken
to equal 12,000 thousands of divine years, each of which equals
360 human years, for a total of 4,320 million human years.
K\d{r}ta, Tret\={a}, Dv\={a}para, and Kali are supposed to last
4,000, 3,000, 2,000, 1,000 divine years respectively. In addition,
there are sandhy\={a}s (twilights) of 800
(two twilights of 400 years), 600, 400, 200 on the
yugas, in order, to give a total span of 12,000 divine years.
Brahm\={a}, the creator of time, is a personification of the 
beginning of the sustaining
principle, to be taken either as Vi\d{s}\d{n}u or \'{S}iva.
Each day of Brahm\={a} is followed by a night of the same duration.
A year of Brahm\={a} equals such 360 day and nights, and the duration
of the universe is the span of 100 Brahm\={a} years.
The largest cycle is 311,040,000 million years. We are supposed to be
in the 55th year of the current Brahm\={a}.
The large cycle is nested in still larger cycles.
Within each kalpa are fourteen secondary cycles, called manvantaras, 
each lasting 306,720,000 years. In each manvantara, humans begin
with a new Manu. We are now in the seventh manvantara of the kalpa,
started by Manu Vaivasvata.

A kalpa equals a thousand mah\={a}yugas, each of which has the
four yugas K\d{r}ta, Tret\={a}, Dv\={a}para, and Kali.
Each manvantara may be divided into 71 mah\={a}yugas.
While the yugas, as defined in the {\em Pur\={a}\d{n}ic} literature
of the 
first millennium CE have extremely large periods in multiples of
the `years of the gods,' it is likely that the four yugas
were originally 4,800, 3,600, 2,400, and 1,200  ordinary years,
respectively.

\section{Astronomy from Lagadha to the Siddh\={a}ntas} 

In this section we review the development
of astronomy between two specific dates, roughly from
1300 BCE to 500 CE.
Although this development is best understood by an examination of 
the Vedic and post-Vedic texts, note that
not all scientific knowledge of those
early times were committed to writing or, if written down, has survived.
There are gaps in the sequence of ideas and these were
filled based on preconceived notions rather than a
sound approach.
New evidence of the past two decades has
contradicted the old 19th-century model of the rise of the
Indian civilization and the new, emerging paradigm has
significant implications for the understanding of the development of
astronomy in India.

Let $R_s$ represent the distance between the earth and the sun,
$R_m$ be the distance between the earth and the moon,
$d_s$ be the diameter of the sun, $d_m$ be the diameter of the
moon, and $d_e$ be the diameter of the earth.
According to {\it PB}, $R_s < 1000~d_e$, and we take that
$R_s \approx 500 d_e$.

It was further known that the
moon and the sun are about 108 times their respective
diameters from the earth.
This could have been easily determined by taking a pole
and removing it to a distance 108 times its height
to confirm that its angular size was equal to that of the sun
or the moon.
Or, we can say that $R_s \approx 108 d_s$ and
$R_m \approx 108 d_m$.

Considering  a uniform  speed  of the sun and the moon and
noting that
the sun completes a circuit in 365.24 days and the
moon 12 circuits in 354.37 days, we find that

\[R_m \approx \frac{354.37 \times 500}{365.24 \times 12} d_e \]
or $R_m \approx 40 d_e$.

By using the relationship on relative sizes that
$R_s \approx 108 d_s \approx 500 d_e$, we know that
$d_s \approx 4.63 \times d_e$.

Assuming that the diameter of the earth was at some time in
the pre-{\em Siddh\={a}ntic} period estimated to be about
900 yojanas, the distance to the moon was then
about 36,000 yojanas and that to the sun about 450,000 yojanas.
It also follows that the relative dimensions of the sun and
the moon were taken to be in the ratio of $12.5:1$.
Knowing that the angular
size of the sun and the moon is about $31.85$ minutes,
the size of the sun is then about 4,170 yojanas
and that of the moon is about 334 yojanas.

A theory on the actual diameters of the
sun, the moon, and the earth indicates a knowledge of eclipses. The
RV 5.40
speaks of a prediction of the duration of a solar eclipse,
so relative fixing of the diameters of the earth, the moon,
and the sun should not come as a surprise.

Also note that the long
periods of Jupiter and Saturn 
require that the sun be much closer to the earth than the
midpoint to the heavens, or push the distance
of the heavens beyond the $1000 d_e$ of {\it PB} and perhaps also
make the distance of the sun somewhat less than $500 d_e$.
We do see these different modifications in the models from
later periods.

The idea that the sun is roughly 500 or so 
earth diameters away from us is much more ancient than Ptolemy
from where it had been assumed to have been borrowed by
the Indians.
This greater antiquity is in accordance with the ideas
of van der Waerden,
who ascribes a primitive epicycle theory to the Pythagoreans.
But it is more likely that the epicycle theory is itself much older than
the Pythagoreans and it is from this earlier source that the later
Greek and Indian modifications to this theory emerged
which explains why the Greek and the Indian models differ in
crucial details.

Did the idea that $R_s \approx 500 d_e$
originate at about the time of
{\it PB}, that is from the second millennium BCE, or is it
older?
Since this notion is in conflict with the data on the periods of
the outer planets, it should predate that knowledge.
If it is accepted that the planet periods were known by the
end of the third millennium BCE, then this 
knowledge must be assigned an even earlier epoch.
Its appearance in {\it PB},
a book dealing primarily
with ritual, must be explained as a remembrance of an old 
idea.
We do know that {\it PB} repeats, almost verbatim, the {\em \d{R}gvedic} account of
a total solar eclipse.

It is certain that the synodic periods were first computed because
the longest period, the 780 days of Mars,
is not too much larger than twice the sun's period.
With Mars as the furthest body in a primitive model,
the sun's distance will have to be reduced to 
about 0.47 of the furthest point.
In order to accommodate the stars, the sun will be brought
even nearer.
When the sidereal orbits of the planets were
understood, sometime in the Vedic period, the space beyond the sun had to be taken
to be vast enough to accommodate the orbits of Jupiter
and Saturn.
The non-circular motions of the planets would require further
changes to the sizes of the orbits and these changes represent
the continuing development of this phase of Indian astronomy.

The
theory that $R_s \approx 500 d_e$ was so strongly entrenched 
that
it became the basis from which different Greek and later Indian 
models emerged.
Ptolemy considers an $R_s$  equal to $600 d_e$ whereas
\={A}ryabha\d{t}a assumes it to be about $438 d_e$.
Thus the Greek and the later Indian modifications to the basic idea
proceeded somewhat differently.

The ideas regarding the distance of the sun hardly changed
until the modern times.
The contradictions in the assumption that the luminaries move with
uniform mean speed and the requirements imposed by the assumed
size of the solar system led to a gradual enlargement of the
models of the universe from about twice that of the distance of the sun in {\it PB}
to one $4.32 \times 10^{6}$ times the distance of the sun by the time of 
\={A}ryabha\d{t}a.
This inflationary model of the universe in
{\it AA} makes a distinction between the distance of the sky (edge
of the planetary system) and
that of the stars which is taken to be a much smaller sixty times the
distance of the sun.
``Beyond the visible universe illuminated
by the sun and limited by the sky is the infinite
invisible universe'' this is stated in a commentary on {\it AA}
by Bh\={a}skara I writing in 629 CE (Shukla, 1976).
The {\em Pur\={a}\d{n}ic} literature,
part of which
is contemporaneous with
\={A}ryabha\d{t}a, reconciles the finite estimates 
of the
visible universe with
the old {\em \d{R}gvedic}
notion of an infinite universe  by postulating the existence of 
an infinite number of universes.

\subsection*{The sizes of the planets}

The ideas on planet sizes can be seen to evolve
from those in the {\em Pur\={a}\d{n}as} to the {\em Siddh\={a}ntas.}
The {\em Pur\={a}\d{n}as} confusingly combine two
different theories, one related to the departure
from the ecliptic by the moon and the other
on the sidereal periods.
The planets are listed in the correct sequence,
supporting the view that the planet periods were
known.
The order of the angular sizes are correctly 
shown as Venus, Jupiter, Saturn, Mars, Mercury
although the fractions stated are not accurate.
Venus and Jupiter are taken to be $\frac{1}{16}$th 
and $\frac{1}{64}$th the size
of the moon whereas the correct fractions are
$\frac{1}{20}$
and $\frac{1}{40}$.
Saturn and Mars were taken to be $\frac{1}{4}$th smaller
than Jupiter and Mercury still smaller by the same fraction
(VaP 53.66-67).

\vspace{0.3in}
Table 3: The planet angular sizes in fractions of the size of the moon

\begin{tabular}{||l|r|r|r||} \hline
Planet & correct size & Pur\={a}\d{n}a & \={A}ryabha\d{t}a\\ \hline
Mercury & $\frac{1}{120}$ & $\frac{1}{112}$& $\frac{1}{15}$\\ \hline
Venus & $\frac{1}{20}$ & $\frac{1}{16}$& $\frac{1}{5}$\\ \hline
Mars & $\frac{1}{100}$ & $\frac{1}{84}$& $\frac{1}{25}$\\ \hline
Jupiter & $\frac{1}{40}$ & $\frac{1}{64}$& $\frac{1}{10}$\\ \hline
Saturn & $\frac{1}{80}$ & $\frac{1}{84}$& $\frac{1}{20}$\\ \hline \hline
\end{tabular}
\vspace{0.3in}

By the time of \={A}ryabha\d{t}a
the {\it relative} sizes of the planets were better estimated
(Table 3). But the angular sizes of the planets are
too large by a factor of 4
excepting Mercury which is too large by a factor of 8.
Overall, the {\em Pur\={a}na} figures are more accurate
and it appears that \={A}ryabha\d{t}a's
overestimation by a factor of 4 may have been
colored by his ideas on optics.

\section{The two halves of the year}
The {\em Br\={a}hma\d{n}as} recognize that the speed
of the sun varies with the seasons.
The year-long rites of the {\em Br\={a}hma\d{n}as} were organized with the
summer solstice ({\it vi\d{s}uvant}) as the middle point.
There were two years: the ritual one started with the
winter solstice ({\it mah\={a}vrata day}), and the civil one
started with the spring equinox ({\it vi\d{s}uva}).
Vedic rites had a correspondence with the different stages of the
year and, therefore, astronomy played a very significant role.
These rites counted the days upto the solstice and in the latter half
of the year, and there is an asymmetry in the two counts.
This is an astronomical parameter, which had hitherto
escaped notice, that allows us to date the
rites to no later than the second millennium BCE.

The {\em Aitareya Br\={a}hma\d{n}a}
4.18 describes how the sun reaches the highest point
on the day called vi\d{s}uvant and how it stays still for a
total of 21 days with the vi\d{s}uvant being the middle day of
this period.
In the {\em Pa\~{n}cavi\d{m}\'{s}a Br\={a}hma\d{n}a}
(Chapters 24 and 25), several year-long rites
are described where the vi\d{s}uvant day is preceded and followed
by three-day periods. This suggests
that the sun was now taken to be more or less still in the heavens
for a total period of 7 days.
So it was clearly understood that the shifting of the rising and
the setting directions had an irregular motion.

\'{S}B 4.6.2 describes the rite
of {\it gav\={a}m ayana}, the ``sun's walk'' or
the ``cows' walk.''
This is a rite which follows the motion of the sun, with its
middle of the vi\d{s}uvant day.

The {\em Yajurveda} (38.20) says that the \={a}havan\={\i}ya or the sky
altar is four-cornered since the sun is four-cornered, meaning
thereby that the motion of the sun is characterized by four
cardinal points: the two solstices and the two equinoxes.

The year-long rites list a total of 180 days before the solstice and
another 180 days following the solstice. 
Since this is reckoning by
solar days, it is not clear stated how
the remaining 4 or 5 days of the year were
assigned.
But this can be easily inferred.

Note that the two basic days in this count are the
vi\d{s}uvant (summer solstice) and the mah\={a}vrata day
(winter solstice)
which precedes it by 181 days in the above counts.
Therefore, even though the count of the latter part of the
year stops with an additional 180 days, it is clear that
one needs another 4 or 5 days to reach the
mah\={a}vrata day in the winter.
This establishes that the division of the year was in the
two halves of 181 and 184 or 185 days.

Corroboration of this is 
suggested by evidence related to an
altar design from the {\em \'{S}atapatha Br\={a}hma\d{n}a}
(\'{S}B 8.6) which is shown in
Figure 4.
This altar represents the path of the sun around the earth.
The middle point, which represents the earth and the
atmosphere is at a slight
offset to the centre.
This fact, and the fact that the number of bricks in the
outer ring are not symmetrically placed, shows that the
four quarters of the year were not taken to be symmetric.

This inequality would have been easy to discover.
The Indians used the reflection of the noon-sun in the water
of a deep well to determine the solstice days.

If one assumes that the two halves of the year are directly in
proportion to the brick counts of 14 and 15 in the two halves
of the ring of the sun, this corresponds to day counts of
176 and 189.
This division appears to have been for the two halves of the
year with respect to the equinoxes if we note that the solstices
divide the year into counts of 181 and 184.

The apparent motion of the sun is the greatest when the
earth is at perihelion and the least when the
earth is at aphelion.
Currently, this speed is greatest in January.
The interval between successive perihelia, the
anomalistic year, is 365.25964 days which is 0.01845 days
longer than the tropical year on which our calendar is based.
In 2000 calendar years, the date of the perihelion advances almost
35 days; in 1000 years, it advances almost a half-year (175 days).
This means that the perihelion movement has a cycle of
about 20000 years.

In the first millennium BCE, the earth was at perihelion
within the interval prior to the winter solstice.
Thus during this period the half of the year from 
the summer solstice to the winter solstice would have been
shorter than the half from the winter solstice to the summer
solstice.
This is just the opposite of what is described in the
rites of the {\em Br\={a}hma\d{n}as}.

It is interesting that the Greeks discovered the asymmetry 
in the quarters of the year about 400 BCE.
Modern calculations show that at this time the four quarters of
the year starting with the winter solstice were 90.4, 94.1, 92.3, and
88.6 days long. The period from the winter solstice to the summer
solstice was then 184.5 days and the perihelion occurred in mid- to
late October.

The count of about 181 days from the winter to the summer solstice
would be true when the perihelion occurs before the summer solstice.
This will require it to move earlier than mid- to late June
and no earlier than mid- to late December. In other words, compared
to 400 BCE, the minimum
number of months prior to October is 4 and the maximum number of
months is 10. This defines periods which are from 6850 years to
17150 years prior to 400 BCE.

These periods appear too early to be considered plausible and this
may reflect the fact that the measurements in those times were not
very accurate.
Nevertheless, it means that the first millennium BCE for the rites
of the {\em Br\={a}hma\d{n}as}
is absolutely impossible.

Since the {\em \'{S}atapatha Br\={a}hma\d{n}a} has lists of teachers that go through
more than fifty generations, we know that the period of the
Br\={a}hma\d{n}as was a long one, perhaps a thousand years.
To be as conservative as possible, one may consider the
period 2000 - 1000 BCE as reasonable for these texts.
The {\em Vedic Sa\d{m}hit\={a}s} are then assigned to the earlier
fourth and third millennia BCE.

\section{The origins of the idea of epicycles}

More than a hundred years ago, Burgess (1860) saw
the Indians as the originators of many of the notions that
led to the Greek astronomical flowering.
This view slowly lost support and then it was believed that
Indian astronomy was essentially derivative and it
owed all its basic ideas to the Babylonians and the Greeks.
It was even claimed that there was no tradition of
reliable observational astronomy in India.

Using statistical analysis of the parameters used
in the many {\em Siddh\={a}ntas}, Billard showed that
the {\em Siddh\={a}ntas} were based on precise observations and
so the theory of no observational tradition in India was
wrong.
This conclusion is reinforced by the fact that
the Vedic books are
according to an astronomical plan.

Earlier, it was believed that
the mah\={a}yuga/kalpa figure of 4,320,000, which occurs in the
{\em Siddh\={a}ntas}, was borrowed from the astronomy$^{12}$ of
the Babylonian Berossos (c. 300 BCE).
But it is more logical to see it derived from the
number 432,000 related to the number of syllables in
the {\em \d{R}gveda} that is mentioned
in the much earlier {\em \'{S}atapatha Br\={a}hma\d{n}a}
(\'{S}B 10.4.2).

The {\em Siddh\={a}ntic} astronomy
has features which are unique to India and it represents
an independent tradition.
In the words of Thurston (1994):

\begin{quote}
Not only did \={A}ryabha\d{t}a believe that the earth rotates,
but there are glimmerings in his system (and other similar
Indian systems) of a possible underlying theory in which the
earth (and the planets) orbits the sun, rather than the sun
orbiting the earth...
The significant evidence comes from the inner planets: the period
of the \'{s}\={\i}ghrocca is the time taken by the planet
to orbit the sun.

\end{quote}

A pure heliocentrism is to be found in the following statement
in the {\em Vi\d{s}\d{n}u Pur\={a}\d{n}a} 2.8:

\begin{quote}
The sun is stationed for all time, in the middle of the day...
The rising and the setting of the sun being perpetually
opposite to each other, people speak of the rising of the
sun where they see it; and, where the sun disappears, there, to 
them, is his setting.
Of the sun, which is always in one and the same place, there
is neither setting nor rising.
\end{quote}

It is not certain that \={A}ryabha\d{t}a was the
originator of the idea of the rotation of the earth.
It appears that the rotation of the earth
is inherent in the notion that the sun never sets that
we find in the {\em Aitareya Br\={a}hma\d{n}a} 2.7:

\begin{quote}
The [sun] never really sets or rises.
In that they think of him ``He is setting,'' having reached the
end of the day, he inverts himself; thus he makes evening below,
day above. Again in that they think of him ``He is rising in the
morning,'' having reached the end of the night he inverts himself;
thus he makes day below, night above. He never sets; indeed he
never sets.
\end{quote}

One way to visualize it is to see the universe as the hollow of
a sphere so that the inversion of the sun now shines the light
on the world above ours. But this is impossible
since the sun does move across the sky during the day and
if the sun doesn't set or rise it doesn't move either.
Clearly, the idea of ``inversion'' denotes nothing but a movement
of the earth.

By examining early Vedic sources
the stages of the
development of the earliest astronomy
become apparent.
After the {\em \d{R}gvedic} stage comes the period 
of the {\em Br\={a}hma\d{n}as} in which we place
Lagadha's astronomy.
The third stage is early {\em Siddh\={a}ntic} and early {\em Pur\={a}\d{n}ic}
astronomy.

These three stages are summarized below:

\begin{enumerate}

\item {\it \d{R}gvedic astronomy (c. 4000? - 2000 BCE)}
Motion of the sun and the moon, nak\d{s}atras, planet periods.
The start of this stage is a matter of surmise but we have
clues such as Vedic myths which have been interpreted
to indicate astronomical events of the fourth millennium
BCE.

\item {\it Astronomy of the Br\={a}hma\d{n}as (2000 - 1000 BCE)}
Astronomy represented by means of geometric altars; non-uniform
motion of the sun and the moon; intercalation for the lunar year;
``strings of wind joined to the sun.''
The {\em Ved\={a}\.{n}ga Jyoti\d{s}a} of Lagadha
must be seen as belonging to the latter part of this stage.
The VJ text that has come down to us appears to be of a later era.
Being the standard manual for determination of the Vedic rites,
Lagadha's work must have served as a ``living'' text where the
language got modified to a later form.

\item {\it Early Siddh\={a}ntic and early Pur\={a}\d{n}ic 
(1000 BCE - 500 CE)}
Here our main sources are the
{\em \'{S}ulbas\={u}tras}, the {\em Mah\={a}bh\={a}rata}, the
early {\em Pur\={a}\d{n}as},
the {\em S\={u}rya Siddh\={a}nta} and other texts.
Further development of the \'{s}\={\i}ghrocca and mandocca cycles, the
concepts of kalpa.
According to tradition, there existed 18 early
{\em Siddh\={a}ntas} composed by S\={u}rya,
Pit\={a}maha, Vy\={a}sa, Vasi\d{s}\d{t}ha, Atri,
Par\={a}\'{s}ara, Ka\'{s}yapa, N\={a}rada,
Garga, Mar\={\i}ci, Manu, A\.{n}giras,
Loma\'{s}a (Romaka), Pauli\'{s}a, Cyavana, 
Yavana, Bh\d{r}gu, and \'{S}aunaka.
Of these, summaries of five are now available in 
the book {\em Pa\~{n}casiddh\={a}ntik\={a}} by
Var\={a}hamihira, and
the {\em S\={u}ryasiddh\={a}nta} has come down in
a later, modified form.

\end{enumerate}

It is significant that the first two stages
and the beginning part of the third stage are well prior to the rise of
mathematical astronomy in Babylonia and in Greece.
The concepts of the 
\'{s}\={\i}ghrocca and mandocca cycles indicate that the motion of
the planets was taken to be fundamentally around the sun, which, in
turn, was taken to go around the earth.

The mandocca, in the case of the sun and the moon, is the apogee where
the angular motion is the slowest and in the case of the other
planets it is the aphelion point of the orbit.
For the superior planets, the \'{s}\={\i}ghrocca  coincides with
the mean place of the sun, and in the case of an inferior planet,
it is an imaginary point moving around the earth with the same
angular velocity as the angular velocity of the planet round the
sun; its direction from the earth is always parallel to the line
joining the sun and the inferior planet.

The mandocca point serves to slow down the motion from the
apogee to the perigee and speed up the motion from the
perigee to the apogee. It is a representation of the non-uniform
motion of the body, and so it can be seen as a direct
development of the idea of the non-uniform motion of
the sun and the moon.

The \'{s}\={\i}ghrocca maps the motion of the planet around the
sun to the corresponding set of points around the earth.
The sun, with its winds that hold the solar system together,
is, in turn, taken to go around the earth.

The antecedents of this system can be seen in the earlier texts.
\'{S}B 4.1.5.16 describes the sun as 
{\it pu\d{s}karam\={a}dityo}, ``the lotus of the sky.''
\'{S}B 8.7.3.10 says:

\begin{quote}
The sun strings these worlds [the earth, the planets,
the atmosphere] to himself on a thread.
This thread is the same as the wind...
\end{quote}

This suggests a central role to the sun in defining the motions
of the planets and ideas such as these must have ultimately
led to the theory of the \'{s}\={\i}ghrocca and the
mandocca cycles.

The theory that the sun was the ``lotus'' [the central point]
of the sky and that it kept the worlds together by its
``strings of wind'' may have given rise to the heliocentric tradition
in India.
The offset of the sun's orbit evolved into the notion of
mandocca and the motions of the planets around the sun were
transferred to the earth's frame through the device of the
\'{s}\={\i}ghrocca.

The {\em Br\={a}hma\d{n}as} consider
non-circular motion of the sun and, by implication, of
the moon and that the sun is taken to be
about 500 earth diameters away from the earth.
Analysis of {\em \d{R}gvedic} astronomy
has shown that planet periods had been determined.
Logically, the next step would be to 
characterize the details of the departure from
the circular motion for the planets.
In VaP 53.71 it is stated that the planets move in
retrograde ({\it vakra}) motion.

Although the extant {\em S\={u}rya Siddh\={a}nta} (SS) is
a late book, it preserves old pre-{\em Siddh\={a}ntic}
ideas on the motions of the planets:

\begin{quote}
Forms of time, of invisible shape, stationed in the
zodiac ({\it bhaga\d{n}a}), called the conjunction
({\it \'{s}\={\i}ghrocca}), apsis ({\it mandocca}),
and node ({\it p\-{a}ta}), are causes of the motion of
the planets.

The planets, attached to these beings by cords of air,
are drawn away by them, with the right and lift hand, forward
and backward, according to nearness, toward their own place.

A wind, called provector ({\it pravaha}) impels them
toward their own apices ({\it ucca}); being drawn away
forward and backward, they proceed by a varying motion.

The so-called apex ({\it ucca}), when in the half-orbit
in front of the planet, draws the planet forward; in like
manner, when in the half-orbit behind the planet, it draws
it backward.

When the planets, drawn away by their apices, move forward
in their orbits, the amount of the motion so caused is
called their excess ({\it dhana}); when they move backward,
it is called their deficiency ({\it \d{r}\d{n}a}). (SS 2.1-5)

\end{quote}

The idea of the sizes was directly related to the
deviation from the ecliptic. The motions were defined
to be of eight different kinds:

\begin{quote}

Owing to the greatness of its orb, the sun is drawn away
only a very little; the moon, by reason of the smallness
of its orb, is drawn away much more;

Mars and the rest, on account of their small size, are,
by the points of focus, called conjunction and apsis, drawn
away very far, being caused to vacillate exceedingly.

Hence the excess and deficiency of these latter is very great,
according to their rate of motion. Thus do the planets, attracted
by those beings, move in the firmament, carried on by the wind.

The motion of the planets is of eight kinds: retrograde ({\it vakra}),
somewhat retrograde ({\it anuvakra}), transverse ({\it ku\d{t}ila}),
slow ({\it manda}), very slow ({\it mandatara}), 
even ({\it sama}), very swift ({\it \'{s}\={\i}ghratara}),
and swift ({\it \'{s}\={\i}ghra}).

Of these, the very 
swift, the swift, the slow, the very slow, and the even
are forms of the motion called direct ({\it \d{r}ju}). (SS 2.9-13)

\end{quote}

\subsection*{The Early Siddh\={a}ntas}
The development of astronomical ideas from
the {\em Ved\={a}\.{n}ga Jyoti\d{s}a} onwards can also
by studied from the information in the Jaina books,
the {\em Mah\={a}bh\={a}rata} and the astronomical
references in the general literature.
For example, the {\em Artha\'{s}\={a}stra} uses a rule for telling
time that is very similar to that in VJ.

The {\em Pa\~{n}casiddh\={a}ntik\={a}}
of
Var\={a}hamihira summarizes five early schools of
{\em Siddh\={a}ntic} astronomy, namely
Pait\={a}maha, V\={a}si\d{s}\d{t}ha,
Romaka, Pauli\'{s}a, and Saura
manly with regard to the calculation of
eclipses.

Owing to the names Romaka and Pauli\'{s}a,
it was assumed that
the PS mostly represents Babylonian and Greek
material.
But such a supposition has no firm evidence to
support to it.
There also exists 
the possibility that an India-inspired
astronomy could have travelled to the West
before the {\em Siddh\={a}ntic} period.

The use of cycles was current during the time
of the {\em \'{S}atapatha Br\={a}hma\d{n}a}.
A modular arithmetic, fundamental to {\em Siddh\={a}ntic}
astronomy, was in use in {\em Ved\={a}\.{n}ga Jyoti\d{s}a}.
The 2,850 
year luni-solar yuga of 
the {\em Romaka Siddh\={a}nta} (PS 1.15)
is derived from the 95-year 
Y\={a}j\~{n}avalkya cycle of the 
{\em \'{S}atapatha Br\={a}hma\d{n}a}, as it is equal to
$30 \times 95$.

Summarizing, the basic features of the
{\em Siddh\={a}ntic} astronomy such as non-circular
orbits of the sun and the moon and the specific notions
of ``ropes of wind'' for the
the planetary system
were already present in the
{\em Br\={a}hma\d{n}as} and they appear in a more
developed form in the primitive epicycle theory
of the {\em S\={u}rya Siddh\={a}nta}.
As the retrograde motions were recognized, the 
orbit sizes were adjusted and made smaller.

\section{Pre-Siddh\={a}ntic Cosmology}

Early texts consider 
light to be like a wind. 
Was any thought 
given to its speed?
Given the nature of the analogy, one would 
expect that this speed was considered finite.
The {\em Pur\={a}\d{n}as} speak of the moving {\it
jyoti\'{s}cakra}, ``the circle of light.''
This analogy or that of the swift arrow let loose from the
bow in these accounts leaves ambiguous whether
the circle of light is the sun or its speeding rays.

We get a specific number 
that could refer to the speed of light
in a late text by
S\={a}ya\d{n}a (c. 1315-1387), prime minister in the court of
Emperors Bukka I of the Vijayanagar Empire and Vedic scholar.
In his commentary on the fourth verse
of the hymn 1.50 of the {\em \d{R}gveda} on the sun, he says
 
\begin{quote}
{\it tath\={a} ca smaryate yojan\={a}n\={a}\d{m} sahasre dve dve \'{s}ate dve ca 
yojane
ekena nimi\d{s}\={a}rdhena kramam\={a}\d{n}a}\\
 
Thus it is remembered:  [O sun] you who
traverse 2,202 {\it yojanas} in half a {\it nime\d{s}a}.
\end{quote}

The same statement occurs in the commentary on the
{\em Taittir\={\i}ya Br\={a}hma\d{n}a} by Bha\d{t}\d{t}a
Bh\={a}skara (10th century?), where it is said to
be an old {\em Pur\={a}\d{n}ic} tradition.
 
The figure could refer to the actual motion of the sun
but, as we will see shortly, that is impossible.
By examining parallels in the {\em Pur\={a}\d{n}ic} literature,
we see it as
an old tradition related to the speed of [sun]light.

The units of {\it yojana} and {\it nime\d{s}a}
are well known. The usual meaning of yojana is about
9.1 miles as in
the {\it Artha\'{s}\={a}stra} where it is
defined as being equal to 8,000 {\it dhanu} or ``bow,'' where
each dhanu is taken to be about
6 feet. \={A}ryabha\d{t}a, Brahmagupta and other astronomers used
smaller yojanas but such exceptional usage was confined to
the astronomers;
we will see that
the {\em Pur\={a}\d{n}as} also use a non-standard measure of
yojana.
As a scholar of the Vedas 
and a non-astronomer, 
S\={a}ya\d{n}a would be expected to use the ``standard''
{\em Artha\'{s}\={a}stra} units.

The measures of time are thus defined in the {\em Pur\={a}\d{n}as}:

\begin{quote}

15 nime\d{s}a = 1 k\={a}\d{s}\d{t}h\={a}

30 k\={a}\d{s}\d{t}h\={a} = 1 kal\={a}

30 kal\={a} = 1 muh\={u}rta

30 muh\={u}rta = 1 day-and-night

\end{quote}

A nime\d{s}a is therefore equal to $\frac{16}{75}$ seconds.

When this statement is
converted into modern units, it does come very close
to the correct figure of 186,000 miles per second!

Such an early knowledge of this number
doesn't sound credible because the speed
of light was determined only in 1675 by Roemer who looked at
the difference in the times that light from Io, one of the moons of
Jupiter, takes to reach the earth based on whether it is on the near side
of Jupiter or the far side.  Until then light was taken to travel with
infinite velocity. 
There is no record of any optical experiments
that could have been performed in India before the modern 
period to measure the speed of light.
 
Maybe S\={a}ya\d{n}a's figure
refers to the speed of the sun in its supposed orbit around the earth.
But that places the orbit of the sun at a distance of over 2,550
million miles.  The correct value is only 93 million miles and until
the time of Roemer the distance to the sun used to be taken to be less
than 4 million miles.  
The Indian astronomical texts place the sun only about half a million
yojanas from the earth.
We show that this figure is connected to {\em Pur\={a}\d{n}ic}
cosmology and, therefore, it belongs, logically, to
the period of pre-{\em Siddh\={a}ntic} astronomy.

\subsection*{Physical ideas in early literature}

The philosophical schools of 
S\={a}\d{m}khya and
Vai\'{s}e\d{s}ika tell us about the old ideas on
light.
According to 
S\={a}\d{m}khya, light is one of the five fundamental
``subtle'' elements ({\it tanm\={a}tra}) out of which
emerge the gross elements.
The atomicity of 
these elements is not specifically mentioned and it appears
that they were actually taken to be continuous.

On the other hand, 
Vai\'{s}e\d{s}ika
is an atomic theory of the physical
world on the nonatomic ground of ether, space and time.
The basic atoms are those of 
earth ({\it p\d{r}thiv\={\i}}), water ({\it \={a}pas}), fire
({\it tejas}), and air ({\it v\={a}yu}), that should
not be confused with the ordinary meaning of these terms.
These atoms are taken to form binary molecules that 
combine further to form larger molecules.
Motion is defined in terms of the movement of the
physical atoms and it appears that it is taken to be
non-instantaneous.  

Light rays are taken to be a stream of
high velocity of tejas atoms.
The particles of light
can exhibit different characteristics
depending on the speed and the
arrangements of the tejas atoms.
 
Although there 
existed several traditions of astronomy in India, only the
mathematical astronomy of the {\em Siddh\={a}ntas} has been
properly examined. 
Some of the information of the non-{\em Siddh\={a}ntic}
astronomical systems is preserved in the
{\em Pur\={a}\d{n}as}.

The {\em Pur\={a}\d{n}ic} astronomy is cryptic, and since
the {\em Pur\={a}\d{n}as} are encyclopaedic texts, with
several layers of writing, presumably by different
authors, there are inconsistencies
in the material.
Sometimes, speculative and the empirical ideas are so
intertwined that without care the material can
appear meaningless.
The {\em Pur\={a}\d{n}ic} geography is quite fanciful and this
finds parallels in its astronomy as well.

We can begin the process of understanding
{\em Pur\={a}\d{n}ic} astronomy by considering its
main features, such as the size of the solar system
and the motion of the sun. 
But before we do so, we will speak briefly of
the notions in the {\em Siddh\={a}ntas}.

\={A}ryabha\d{t}a 
in his {\it \={A}ryabha\d{t}\={\i}ya (AA)}
deals with the question of the size of the universe.
He defines a
{\it yojana} to be
8,000 {\it n\d{r}},
where a {\it n\d{r}} is the height of a man;
this makes his yojana ($y_a$) approximately 7.5 miles.
Or $y_s \approx \frac{6}{5} y_a$,
where $y_s$ is the standard {\em Artha\'{s}\={a}stra}
yojana.
AA 1.6
states that the orbit of the sun is
2,887,666.8 {\it yojanas} and that of the
sky is 12,474,720,576,000 {\it yojanas}.

There is no mention by \={A}ryabha\d{t}a of a
speed of light. But 
the range of light particles is taken to be finite,
so it must have been assumed that the particles in the
``observational universe'' do not penetrate to
the regions beyond the ``orbit of the sky.''
This must have been seen in the 
analogy of the gravitational pull
of the matter just as other particles fall back
on the earth after reaching a certain height.

The orbit of the sky is $4.32 \times 10^6$ greater than the
orbit of the sun.
It is clear
that this enlargement was inspired by cosmological ideas.

The diameters of the earth, the sun, and the moon are taken to
be 1,050, 4,410 and 315 yojanas, respectively.
Furthermore, {\it AA} 1.6 implies
the distance to the sun, $R_s$, 
to be 459,585 yojanas, and that to the moon, $R_m$, as 34,377 yojanas.
These distances are in the correct proportion related to
their assumed sizes given that the distances are
approximately 108 times the corresponding diameters.

Converted to the standard {\it Artha\'{s}\={a}stra} units, the diameters
of the earth and the sun are about 875 and 3,675 yojanas,
and the distance to the sun
is around 0.383 million yojanas.

\subsection*{Pur\={a}\d{n}ic cosmology}

The {\em Pur\={a}\d{n}ic} material is
closer to the knowledge
of the Vedic times.
Here we specifically consider 
the {\em V\={a}yu Pur\={a}\d{n}a} (VaP), 
{\em Vi\d{s}\d{n}u Pur\={a}\d{n}a} (ViP),
and {\em Matsya Pur\={a}\d{n}a} (MP).
VaP and ViP are generally believed to be amongst the earliest
{\em Pur\={a}\d{n}as} and at least 1,500 years old.
Their astronomy is prior to the {\em Siddh\={a}ntic}
astronomy of \={A}ryabha\d{t}a and his successors.

The {\em Pur\={a}\d{n}as} instruct through myths and 
this mythmaking can be seen in their approach to
astronomy also.
For example, they speak of seven underground
worlds
below the orbital plane of the planets
and of seven ``continents'' encircling the earth.
One has to take care to separate this
imagery, that parallels the conception
of the seven centres of the human's
psycho-somatic body, from the underlying cosmology 
of the {\em Pur\={a}\d{n}as}
that is their primary concern in their {\it jyoti\d{s}a}
chapters.
The idea of seven regions of
the universe is present in the {\em \d{R}gveda} 1.22.16-21 where
the sun's stride is described as {\it saptadh\={a}man},
or taking place in seven regions.

The different {\em Pur\={a}\d{n}as} appear to 
reproduce the same cosmological material.
There are some minor differences in 
figures that may be a result of wrong 
copying by scribes who did not understand the
material.
Here we mainly follow ViP.

ViP 2.8 describes the sun to be 9,000 yojanas in
length and to be connected by an axle that is
$15.7 \times 10^6$ yojanas long to the 
M\={a}nasa mountain and another axle
45,500 yojanas long connected to the pole star.
The distance of 15.7 million yojanas between the
earth and the sun is much greater than the
distance of 0.38 or 0.4375 million yojanas that we
find in the {\em Siddh\={a}ntas} and other early books.
This greater distance is stated without a corresponding change in
the diameter of the sun.

Elsewhere, in VaP 50, it is stated that the sun
covers 3.15 million yojanas in a muh\={u}rta.
This means that the distance covered in a day are 94.5
million yojanas.
MP 124 gives the same figure.
This is in agreement with the view that the sun is
15.7 million yojanas away from the earth.
The specific speed given here, translates to 
116.67 yojanas per half-nime\d{s}a.

The size of the universe is described in two different
ways, through the ``island-continents'' and through
heavenly bodies.
The geography of the {\em Pur\={a}\d{n}as} describes a central
continent, Jambu, surrounded by alternating bands of ocean and land.
The seven island-continents 
of Jambu, Plak\d{s}a, \'{S}\={a}lmala, Ku\'{s}a,
Kraunca, \'{S}\={a}ka, and Pu\d{s}kara are encompassed, successively,
by seven oceans; and each ocean and continent is, respectively,
of twice the extent of that which precedes it.
The universe is seen as a sphere of size 500 million yojanas.

The continents are
imaginary regions and they should not be
confused with the continents on the earth.
Only certain part of the innermost planet, Jambu, that deal with
India have
parallels with
real geography.

The inner continent is taken to be 
16,000 yojanas as the base of the world axis.
In opposition to the interpretation by
earlier commentators, who took the
increase in dimension by a factor of
two is only across the seven ``continents,''
we take it to apply to the ``oceans'' as well. 
At the end of the seven island-continents is a region that is twice the preceding
region. Further on, is the Lok\={a}loka mountain, 10,000 yojanas in breadth,
that marks the end of our universe.

Assume that the size of the Jambu is $J$ yojana,
then the size of the universe is:

\begin{equation}
U = J ( 1 +2 +2^2+ 2^3 +2^4 +2^5 +2^6 +2^7 +2^8 +2^9 +2^{10} +2^{11} +2^{12} +2^{13} +2^{14}) +20,000 
\end{equation}

Or,

\begin{equation}
U =  32,767 J + 20,000~ yojanas
\end{equation}

If U is 500 million yojanas, then J should be about 15,260 yojanas.
The round figure of 16,000 is mentioned as the width of the base of
the Meru, the world axis, at the surface of the earth.
This appears to support our interpretation.

Note that the whole description of the {\em Pur\={a}\d{n}ic} cosmology
had been thought to be inconsistent because 
an erroneous interpretation of the increase in the sizes of
the ``continents'' had been used.

When considered in juxtaposition with the preceding numbers,
the geography of concentric continents is
a representation of the plane of the earth's rotation, with
each new continent as the orbit of the next ``planet''.

The planetary model in the {\em Pur\={a}\d{n}as} is
different from that in the {\em Siddh\={a}ntas}.
Here the moon as well as the planets are in orbits
higher than the sun. 
Originally, this supposition for the moon may
have represented the fact that it goes higher than the
sun in its orbit.
Given that the moon's inclination is $5^\circ$ to
the ecliptic, its declination can be $28.5^\circ$
compared to the sun's maximum declination of
$\pm 23.5^\circ$.
This ``higher'' position must have been, at
some stage, represented literally by a
higher orbit. To make sense with the
observational reality, it became necessary for
the moon is taken to be twice as large as
the sun. 
That this is a jumbling up of two different theories
is clear from the fact that the planets are listed
in the correct sequence determined by their
sidereal periods.

The distances of the planetary orbits beyond the sun are
as follows:

\vspace{0.3in}
Table 4: From the earth to the Pole-star

\begin{tabular}{||l|r||} \hline
Interval I & yojanas\\ \hline
Earth to Sun & 15,700,000\\
Sun to Moon  & 100,000 \\
Moon to Asterisms  & 100,000 \\
Asterisms to Mercury  & 200,000 \\
Mercury to Venus  & 200,000 \\
Venus to Mars  & 200,000 \\
Mars to Jupiter  & 200,000 \\
Jupiter to Saturn  & 200,000 \\
Saturn to Ursa Major  & 100,000 \\
Ursa Major to Pole-star  & 100,000 \\\hline
Sub-total  & 17,100,000 \\ \hline
\end{tabular}
\vspace{0.3in}

Further spheres are postulated 
beyond the pole-star.
These are the Maharloka, the Janaloka, the Tapoloka, and the
Satyaloka. Their distances are as follows:

\vspace{0.3in}
Table 5: From Pole-star to Satyaloka

\begin{tabular}{||l|r||} \hline
Interval II & yojanas\\ \hline
Pole-star to Maharloka  & 10,000,000 \\
Maharloka to Janaloka  & 20,000,000 \\
Janaloka to Tapoloka  & 40,000,000 \\
Tapoloka to Satyaloka  & 120,000,000 \\\hline
Grand Total  & 207,100,000 \\ \hline
\end{tabular}

\vspace{0.3in}

Since the last figure is the distance from the earth, the
total diameter of the universe is 414.2 million yojanas,
not including the dimensions of the various heavenly bodies
and {\it lokas}.
The inclusion of these may be expected to bring this
calculation in line with the figure of 500 million
yojanas mentioned earlier.

Beyond the universe lies the limitless {\it pradh\={a}na}
that has within it countless other universes.

{\em Pur\={a}\d{n}ic} cosmology views the universe as going
through cycles of creation and destruction of 8.64 billion years.
The consideration of a universe of enormous size must have
been inspired by a supposition of enormous age.

\subsection*{Reconciling Pur\={a}\d{n}ic and Standard Yojanas}

It is clear that the {\em Pur\={a}\d{n}ic} yojana ($y_p$) are
different from the {\em Artha\'{s}\={a}stra} yojana ($y_s$).
To find the conversion factor, we equate the distances to
the sun.

\begin{equation}
0.4375 \times 10^6 ~y_s = 15.7 \times 10^6 ~y_p
\end{equation}

In other words,

\begin{equation}
1 ~y_s \approx 36 ~y_p
\end{equation}

The diameter of the earth should now be about
$875 \times 36 \approx 31,500 ~ y_p$.
Perhaps, this was taken to be 32,000 $y_p$,
twice the size of Meru.
This understanding is confirmed by the statements in
the {\em Pur\={a}\d{n}as}.
For example, MP 126 says that 
the size of Bh\={a}ratavar\d{s}a (India)
is 9,000 $y_p$, 
which is roughly correct.

We conclude that the kernel of the {\em Pur\={a}\d{n}ic} system
is consistent with the {\em Siddh\={a}ntas}.
The misunderstanding of it arose because attention was
not paid to their different units of distance.

\subsection*{Speed of the sun}

Now that we have a {\em Pur\={a}\d{n}ic} context,
the statement that the sun has  the speed
of 4,404 
yojanas per nime\d{s}a can be examined.

We cannot be absolutely certain what yojanas did S\={a}ya\d{n}a have in mind:
standard, or {\em Pur\={a}\d{n}ic}.
But either way it is clear from the summary of {\em Pur\={a}\d{n}ic}
cosmology that this speed could not be the speed of the sun.
At the distance of 15.7 million yojanas, the sun's speed 
is only 121.78 yojanas ($y_p$) per half-nime\d{s}a.
Or if we use the the figure from VaP,
it is 116.67.
Converted into the standard yojanas, this number
is only 3.24 $y_s$ per half-nime\d{s}a.

S\={a}ya\d{n}a's speed is about 18 times
greater than the supposed speed of the sun in $y_p$
and $2 \times 18^2$ greater than the speed in $y_s$.
So either way, a larger number with a definite
relationship to the actual speed of the sun was chosen
for the speed of light.

The {\em Pur\={a}\d{n}ic} size of the universe is 13 to 16
times greater than the orbit of the sun, not counting
the actual sizes of the various heavenly bodies.
Perhaps, the size was taken to be 18 times greater
than the sun's orbit.
It seems reasonable to assume, then, that if the
radius of the universe was taken to be about 282 million
yojanas, a speed was postulated for light so that
it could circle the farthest path in the universe within
one day.
This was the physical principle at the basis of
the {\em Pur\={a}\d{n}ic} cosmology.

We have seen that the astronomical 
numbers in the {\em Pur\={a}\d{n}as} are much more consistent 
amongst themselves, and with
the generally accepted 
sizes of the solar orbit, than has been hitherto assumed.
The {\em Pur\={a}\d{n}ic} geography must not be taken
literally.

We have also shown that
the S\={a}ya\d{n}a's figure of 2,202 yojanas
per half-nime\d{s}a is consistent with
{\em Pur\={a}\d{n}ic} cosmology where the
size of ``our universe'' is a function of
the speed of light.
This size represents the space that can be
spanned by light in one day.

It is quite certain that the figure for speed was obtained either by
this argument or it was obtained by taking the
postulated speed of the sun in the {\em Pur\={a}\d{n}as} and
multiplying that by 18, or by multiplying the
speed in standard yojanas by $2 \times 18^2$.
We do know that 18 is a sacred number in the {\em Pur\={a}\d{n}as},
and the fact that multiplication with this special number gave
a figure that was in accord with the spanning of light in the
universe in one day must have given it a special significance.

Is it possible that the number 2,202 arose because of a mistake
of multiplication by 18 rather than a corresponding division
(by 36) to reduce the sun speed to standard yojanas? The answer
to that must be ``no'' because such a mistake is too egregious.
Furthermore, S\={a}ya\d{n}a's own brother M\={a}dhava was a
distinguished astronomer and the incorrectness of this figure
for the accepted speed of the sun would have been obvious to him.

If S\={a}ya\d{n}a's figure was derived from a postulated size of the 
universe, how was that huge size, so central
to all Indian thought, arrived at?
A possible explanation is that the
physical size of the universe was taken to parallel
the estimates of its age.
These age-estimates were made larger and larger to 
postulate a time when the periods of all the
heavenly bodies were synchronized.
The great numbers in the {\em Pur\={a}\d{n}as} suggest
that the concepts of mah\={a}yuga and
kalpa
must have had an old
pedigree and they can be viewed as generalizations
of the notion of yuga.

The speed of light was
taken to be $2 \times 18^2$ greater than the speed of
the sun in standard yojanas so that
light can travel the entire postulated size of the universe
in one day. It is a lucky chance that
the final number turned out to be exactly equal to
the true speed.
This speed of light must be considered
the most astonishing ``blind hit'' in the history of science!
But it is consistent with {\em Pur\={a}\d{n}ic} model of
the cosmos and it is, in most likelihood, a
pre-\={A}ryabha\d{t}a figure.

\section{The Later Siddh\={a}ntic Period}

This period begins with \={A}ryabha\d{t}a (born 476 CE) who established
two systems by his works 
{\em \={A}ryabha\d{t}\={\i}ya} and
{\em \={A}ryabha\d{t}asiddh\={a}nta}. Of these, the first
has exercised great influence, especially in South India, while
the second is lost but elements of it are known through quotations in other
texts and in criticism.
These two constitute the {\em \={A}ryapak\d{s}a} and the
{\em \={A}rdhar\={a}trikapak\d{s}a}, respectively.
The {\em S\={u}ryasiddh\={a}nta}, which has come down to us in
a later recension, is based primarily according to
{\em \={A}rdhar\={a}trikapak\d{s}a}.

The {\em Br\={a}hma-sphu\d{t}a-siddh\={a}nta} of
Brahmagupta (born 598 CE) has been extremely influential in north
and west India and in the Arabic world, through its translation 
called {\em Sindhind}.
A rival to the \={A}ryabha\d{t}a systems, this is
also called the {\em Brahmapak\d{s}a}.

The later improvements to these {\em Siddh\={a}ntas} required
{\em b\={\i}ja}-corrections.
A later text by Lalla (8th-9th cent.) is based on the
{\em \={A}rdhar\={a}trikapak\d{s}a} but 
it also incorporates ideas from the {\em Brahmapak\d{s}a}.

The {\em Siddh\={a}nta-\'{S}iroma\d{n}i} of Bh\={a}skara II (c.1150)
is the most comprehensive of the Indian {\em Siddh\={a}ntas.}
It is based on the Brahmapak\d{s}a.
The epicyclic-eccentric theories are further developed
to account for the motions of the planets.

N\={\i}lakan\d{t}ha Somay\={a}ji (1444-1550) corrects the
{\em \={A}ryapak\d{s}a} constants in his
{\it Siddh\={a}nta-darpa\d{n}a} and
{\it Tantrasa\d{m}graha}.
In a recent review (Ramasubramanian et al, 1994), it has been
argued that N\={\i}lakan\d{t}ha's revision of the
planetary model for the interior planets Mercury and
Venus led to a better equation of the center for
``these planets than was available either in the earlier
Indian works or in the Islamic or European traditions
of astronomy till the work of Kepler, which was to come more
than a hundred years later.''

\section{Concluding Remarks}

This essay summarizes the essential points of the
emerging new understanding of the rise and
development of Indian astronomy. We have traced
the gradual development of many ideas of
astronomy in the pre-{\em Siddh\={a}ntic} period, but we are
not in a position to completely discount all
outside influences especially because
considerable interaction existed in the
ancient world and ideas must have travelled in several
directions.

Further evidence in support of the flow of
ideas from India to the West has recently become clear.
Recent studies of Celtic material indicate
that a calendar similar to the 5-year yuga
of VJ with two intercalary months was current amongst the Druids
(Ellis, 1994:230-231).
The connections between the Vedic and
the Druidic material must predate the
rise of the astronomy in Mesopotamia, because
otherwise the more direct Mesopotamian theories
would have won out against the complex Vedic system.
The Druids also appear to have counted in months
of 27 days similar to the conjoining of the moon with
the nak\d{s}atras of Vedic astronomy.
This supports the idea of transmission from India into
Europe.
This idea is further supported by a new analysis of
a {\em \d{R}gvedic} hymn on Vena (Kak, 1998c) which suggests that
the seed ideas of
the Venus mythologies of the Mesopotamians, the Greeks,
and the later {\em Pur\={a}\d{n}ic} period are all
present in the Vedic texts.

A figure from the neolithic/chalcolithic period
of Indian art (5000 BCE ?) (Kak, 1998a) appears to
be the prototype of the ``Gilgamesh'' or ``hero'' motif with a god or
goddess holding
back two beasts on either side.
The beasts are without their front ends, so clearly
the depiction is symbolic.
David Napier (1986) has argued for a transmission of Indian motifs
into Greece in the second millennium BCE.
As another example consider the Gundestrup cauldron,
found in Denmark a hundred years ago. This silver bowl
has been dated to around the middle of the 2nd century
BCE.
The iconography is strikingly Indic as
clear from the elephant (totally out of
context in Europe) with the goddess and
the yogic figure (Taylor, 1992).
The unicorn figure of European mythology appears to be
based on the Indian conception of the unicorn seen
in the many fine representations of it in Harappan art
and also its celebration in the Vedic and Pur\={a}\d{n}ic
texts as eka\'{s}\d{r}\.{n}ga.

According to Seidenberg's analysis (Seidenberg, 1962, 1978) Indian geometry and mathematics
predate Babylonian and Greek mathematics.
It is likely that the cultural processes that were
responsible for the spread of Indian mathematics and art were also
responsible for the spread of Indian astronomy
during the pre-{\em Siddh\={a}ntic} period.
Doubtless, there existed transmission of Western (Babylonian and Greek) ideas
to India as well.

The most important conclusion of the new findings is that
there existed a much greater traffic of ideas
in all directions in the ancient world than has hitherto been
supposed.
\section*{Abbreviations}
\begin{tabular}{ll}
AA &	Aitareya \={A}ra\d{n}yaka\\
AB &	 Aitareya Br\={a}hma\d{n}a\\
ABA &	\={A}ryabha\d{t}\={\i}ya of \={A}ryabha\d{t}a\\
ASS &	 \={A}pastamba \'{S}ulbas\={u}tra\\
AV &	 Atharvaveda\\
BSS &	 Baudh\={a}yana \'{S}ulbas\={u}tra\\
BU &	 B\d{r}had\={a}ra\d{n}yaka Upani\d{s}ad\\
CU &	 Ch\={a}ndogya Upani\d{s}ad\\
KB &	Kau\d{s}\={\i}taki Brahma\d{n}a\\
MP &    Matsya Pur\={a}\d{n}a\\
PB &	Pa\~{n}cavi\d{m}\'{s}a Brahma\d{n}a\\
PS &	 Pa\~{n}casiddh\={a}ntik\={a}\\
RV &	 \d{R}gveda\\
\'{S}B &	 \'{S}atapatha Br\={a}hma\d{n}a\\
SS &	S\={u}rya Siddh\={a}nta\\
\'{S}U & \'{S}vet\={a}\'{s}vatara Upani\d{s}ad \\
TB &	 Taittir\={\i}ya Br\={a}hma\d{n}a \\
TS &	 Taittir\={\i}ya Sa\d{m}hit\={a}\\
VaP &    V\={a}yu Pur\={a}\d{n}a\\
ViP &    Vi\d{s}\d{n}u Pur\={a}\d{n}a\\
VJ &	 Ved\={a}\.{n}ga Jyoti\d{s}a\\
\end{tabular}

\section*{Bibliography}
\begin{description}
 
\item
Aaboe, A. Scientific astronomy in antiquity. {\em  Phil, Trans. Roy. Soc.
Lond. A} 276: 21-42, 1974.

\item
Ashfaque, S.M. ``Primitive astronomy in the Indus
civilization.' In {\it Old Problems and New Perspectives in the
Archaeology of South Asia,} ed. J.M. Kenoyer, 207-215, Madison, 1992.

\item Billard, Roger. {\it L'astronomie Indienne.} Paris:
'{E}cole fran\c{c}aise d'Extr\^{e}me-Orient, 1971.

\item Burgess, E. (tr.) {\it The S\={u}rya Siddh\={a}nta}.
Delhi: Motilal Banarsidass, 1989 (1860).

\item Ellis, Peter B. {\em The Druids.} London: Constable \& Company, 1994.
 
\item Feuerstein, G.,  S. Kak and D. Frawley. {\it In Search of the
Cradle of Civilization.} Wheaton: Quest Books, 1995.
 
\item Francfort, H.-P. Evidence for Harappan irrigation system in
Haryana and Rajasthan. {\it Eastern Anthropologist} 45: 87-103, 1992.
 
\item Frawley, David. ``Planets in the Vedic literature.''
{\it Indian Journal of History of Science.} 29: 495-506, 1994.

\item Issue devoted to
history of Indian astronomy. {\em Indian Journal of History of Science}, 20, 1985.

\item Kak, Subhash. ``Astronomy in the Vedic altars and the \d{R}gveda.''  {\em Mankind
Quarterly} 33: 43-55, 1992.
 
\item Kak, Subhash, and David Frawley. ``Further observations on the Rigvedic code.''  
{\em Mankind Quarterly} 33: 163-170, 1992.
 
\item Kak, Subhash. ``Astronomy in \'{S}atapatha Br\={a}hma\d{n}a.''
{\em Indian Journal of History of Science} 28: 15-34, 1993.
 
\item Kak, Subhash. ``Astronomy of the Vedic altars.''  {\em Vistas in Astronomy} 
36: 117-140, 1993.
 
\item Kak, Subhash. ``The structure of the \d{R}gveda.'' {\em Indian 
Journal of History of Science} 28: 71-79, 1993.
 
\item Kak, Subhash. ``Planetary periods from the Rigvedic code.''  {\em Mankind
Quarterly} 33: 433-442, 1993.

\item Kak, Subhash. ``The Astronomical Code of the Rigveda.'' {\em Puratattva: 
Bulletin of the
Indian Archaeological Society} 25: 1-30, 1994/5.
 
\item Kak, Subhash. {\it The Astronomical Code of the \d{R}gveda.}
New Delhi: Aditya, 1994.

\item Kak, Subhash. ``The astronomy of the age of geometric altars.''
{\it Quarterly Journal of the Royal Astronomical Society} 36: 385-396, 1995.
 
\item Kak, Subhash. ``From Vedic science to Ved\={a}nta.''
{\it The Adyar Library Bulletin} 59: 1-36, 1995.
 
\item Kak, Subhash. ``Knowledge of planets in the third millennium B.C.''
{\it Quarterly Journal of the Royal Astronomical Society.} 37: 709-715, 1996.
 
\item Kak, Subhash. ``Archaeoastronomy and literature.''
{\it Current Science} 73: 624-627, 1997.

\item Kak, Subhash. `Mind, immortality and art.' Presented at 
the {\it International Seminar on {\em Mind, Man and Mask},
Indira Gandhi National Centre for the Arts, New Delhi, Feb 24-28, 1998a}.

\item Kak, Subhash. ``Astronomy and its role in Vedic culture.'' Chapter 23 in
{\it Science and Civilization in India, Vol. 1,
The Dawn of Indian Civilization, Part 1}, edited by G.C. Pande,
Oxford University Press, Delhi, 1998b, pp. 507-524.

\item Kak, Subhash. ``Vena, Veda, Venus.'' {\it Indian Journal of History
of Science} 33: 25-30, 1998c.

\item Kak, Subhash. ``S\={a}ya\d{n}a's astronomy.'' {\em Indian Journal of History of
Science} 33: 31-36, 1998d.

\item Kak, Subhash. ``Early theories on the distance to the sun.'{\it
Indian Journal of History of Science} 33: 93-100, 1998e.

\item Kak, Subhash. ``The orbit of the sun in the Br\={a}hma\d{n}as.'' {\it
Indian Journal of History of Science} 33: 175-191, 1998f.

\item Lal, B.B. {\it The Earliest Civilization of South Asia.}
New Delhi: Aryan International, 1997.

\item Napier, A. David. {\it Masks, Transformation, and Paradox.}
Berkeley: University of California Press, 1986.

\item Pandya, H., S. Dikshit, and M.N. Kansara, eds.
{\em Issues in Vedic Astronomy and Astrology.}
New Delhi: Rashtriya Veda Vidya Pratishthan, 1992.

\item Pingree, David. ``The recovery of early Greek astronomy from
India.'' {\it Journal for the History of Astronomy} 7:
109-123, 1976.

\item Pingree, David. ``Two treatises on Indian astronomy.''
{\it Journal for the History of Astronomy} 11:
58-62, 1980.

\item Pingree, David. ``History of mathematical astronomy in India.''
In {\em Dictionary of Scientific Biography}, C.S. Gillespie, ed., 533-633.
New York: Charles Scribner's Sons, 1981.

\item 
Ramasubramanian, K., M.D. Srinivas, and M.S. Sriram. ``Modification of
the earlier Indian planetary theory by the Kerala astronomers
(c. 1500 AD) and the implied heliocentric picture of planetary 
motion.'' {\em Current Science} 66: 784-790, 1994.

\item de Santillana, G. and von Dechend, H. {\it Hamlet's Mill: An Essay
on Myth and the Frame of Time.} Gambit, Boston, 1969.

\item Sarma, K.V. {\em A History of the Kerala School of 
Hindu Astronomy.}
Hoshiarpur: Vishveshvaranand Institute, 1972.

\item Sarma, K.V. ``A survey of source materials.''
{\em Indian Journal of History of Science} 20: 1-20, 1985.

\item Sastry, T.S. Kuppanna. {\em Ved\={a}\.{n}ga Jyoti\d{s}a of Lagadha.}
New Delhi: Indian National Science Academy, 1985.

\item Seidenberg, A. ``The origin of geometry.''
{\it Archive for History of Exact Sciences.} 1: 488-527, 1962.

\item Seidenberg, A. ``The origin of mathematics.''
{\it Archive for History of Exact Sciences.} 18: 301-342, 1978.
 
\item Sengupta, P.C. {\it Ancient Indian Chronology.}
Calcutta: University of Calcutta Press, 1947.
 
\item Shaffer, J. and D.L. Lichtenstein. ``The concept of cultural
tradition and paleoethnicity in South Asian archaeology.''
In {\it The Indo-Aryans of
South Asia,} G. Erdosy. ed. Berlin: Walter de Gruyter, 1995.
 
\item Shukla, Kripa Shankar. {\em \={A}ryabha\d{t}a: Indian
Mathematician and Astronomer.} New Delhi: Indian National Science
Academy, 1976.

\item Taylor, Timothy. ``The Gundestrup cauldron.'' {\it Scientific American}
266 (3), 84-89, 1992.

\item Thurston, H., {\it Early Astronomy}.
New York: Springer-Verlag, 1994.

\item 
Vidy\={a}la\.{n}k\={a}ra, V.
{\em \'{S}atapatha Br\={a}hma\d{n}astha Agnicayana Sam\={\i}k\d{s}\={a}.}
Bahalgarh, 1985.

\item  van der Waerden, B.L. ``The earliest form of the epicycle
theory.'' {\it Journal for the History of Astronomy} 5:
175-185, 1974.

\item Wakankar, Vishnu S. ``Rock painting in India.'' In
{\it Rock Art in the Old World,} ed. M. Lorblanchet, 319-336.
New Delhi, 1992.

\end{description}

\newpage

\subsection{Figure captions}

\indent

Figure 1. The falcon altar which represented the days in the year

Figure 2. The traditional representation of the nak\d{s}atras

Figure 3. The Indian zodaic along with the signs for the sun, five planets,
the moon and its ascending and descending nodes

Figure 4. The asymmetric orbit of the sun around the earth

\end{document}